\documentclass{pasa}%

\usepackage{graphicx}
\usepackage{rotating}
\usepackage{tabularx}
 \usepackage{proofread}
    \usepackage[utf8]{inputenc}

\title[Comparison of Gaia and Hipparcos parallaxes of CVBS]{Comparison of Gaia and Hipparcos parallaxes of close visual binary stars and the impact on determinations of their masses}

\author[M. A. Al-Wardat et al.]{
M. A. Al-Wardat,$^{1,2}$\thanks{E-mail: malwardat@sharjah.ac.ae}
A. M. Hussein,$^{3}$
H. M. Al-Naimiy $^{1,2}$
and M. A. Barstow$^{4}$\\
$^{1}$Department of Applied Physics and Astronomy, College of Sciences, University of Sharjah, Sharjah 27272, UAE\\
$^{2}$Sharjah Academy for Astronomy, Space Sciences and Technology, University of Sharjah, Sharjah 27272, UAE\\
$^{3}$Department of Physics, Al al-Bayt University, Mafraq 25113, Jordan\\
$^{4}$School of Physics and Astronomy, University of Leicester, University Road, Leicester, LE1 7RH, United Kingdom}%

\jid{PASA}
\doi{\href{https://www.cambridge.org/core/journals/publications-of-the-astronomical-society-of-australia/article/abs/comparison-of-gaia-and-hipparcos-parallaxes-of-close-visual-binary-stars-and-the-impact-on-determinations-of-their-masses/2CF6E59430E77399994C6F4F8838745B}{10.1017/pas.\the\year}. The paper has been accepted for publication in PASA, 16th Nov. 2020}
\jyear{\the\year}

\usepackage{aas_macros}
\usepackage{hyperref}
\hypersetup{colorlinks,citecolor=blue,linkcolor=blue,urlcolor=blue}
\begin{document}

\begin{frontmatter}
\maketitle

\begin{abstract}
Precise measurement of the fundamental parameters of stellar systems, including mass and radius, depends critically on how well the stellar distances are known. Astrometry from space provides parallax measurements of unprecented accuracy, from which distances can be derived, initially from the Hipparcos mission, with a further refinement of that analysis provided by van Leeuwen in 2007. The publication of the Gaia DR2 catalogue promises a dramatic improvement in the available data. We have recalculated the dynamical masses of a sample of 1700 close visual binary stars using Gaia DR2 and compared the results with masses derived from both the original and enhanced Hipparcos data. We show the Van Leeuwen analysis yields results close to those of Gaia DR2, but the latter are significantly more accurate. We consider the impact of the Gaia DR2 parallaxes on our understanding of the sample of visual binaries.
\end{abstract}

\end{frontmatter}

\section{INTRODUCTION }
\label{sec:intro}

The Hipparcos satellite was the first space mission to be devoted to astrometry.  The European Space Agency (ESA) launched the telescope in 1989, which successfully accomplished its mission during the period from 1991 to 1993, making measurements of nearly 118,000 stars. The results were published in 1997 yielding the Hipparcos and Tycho Catalogues ~\citep{1997yCat.1239....0E}. In 2007,  F. van Leeuwen, a member of the Hipparcos team, published a new reduction of  the Hipparcos catalogue that improved the accuracy of the measured parameters ~\citep{2007AA...474..653V}.
%But, tell Gaia sent its first results, there was no way to judge van Leeuwen's work, whether he was right or not!.

%Since parallaxes of multiple stellar systems or even single stars are crucial in identifying their physical and geometrical parameters, it is important to  to use different ways of measuring or calculating these parallaxes as a double check.

Launched in 2013, the ESA Gaia mission obtained precise astrometry and photometry for approximately 1.7 billion stars, a 10,000 fold increase on Hipparcos. The first data set was published in 2016 as Data Release 1 (DR1) ~\citep{2016A&A...595A...1G}, and the second data set was published in 2018 ~\citep{2018yCat.1345....0G, 2018A&A...616A...2L}.
 Gaia DR2 is already revolutionizing many areas of stellar astrophysics. Many researchers now rely on and confirm the advantages of Gaia measurements (e.g.~\cite{2018ApJ...862...61S}).

Stellar binary systems are a key source of physical parameters of individual stars, especially masses, radii and distances. However, the  fact of binarity can make parallax measurements difficult and affect their accuracy. For example, ~\cite{1998AstL...24..673S} pointed this out, noting that Hipparcos parallax measurements of binary and multiple systems are, in some cases, distorted by the orbital motion of the components of such systems. To study parallax measurements of stars in binaries and assess the relative reliability of Hipparcos,  van Leeuwen and Gaia DR2 we selected a sample of  close visual binary stars (CVBSs) taken from the Sixth Catalog of Orbits of Visual Binary Stars (ORB6) ~\citep{2001AJ....122.3480H}. This is a useful reference catalogue and data set as it includes around 2,900 solved orbits of approximately 2,700 CVBSs distributed all over the celestial sphere. Among the stars listed in the sixth catalog, we found 1700 CVBSs with parallax measurements given by Hipparcos 1997, van Leeuwen reduced Hipparcos 2007, and Gaia DR2 2018.
With parallax errors improved over both the original Hipparcos catalogue and the re-analysis of \citep{2007AA...474..653V}, the Gaia DR2 catalogue and the newly computed masses and radii should supersede any earlier results. Nevertheless, a comparison between DR2 and the earlier parallax measurements provides a useful benchmark test for the latest results.

\begin{sidewaystable*}
	\centering
	\caption{Fundamental, orbital and observational trigonometric parallax data,first twenty-five lines.}
	\label{orb}
	\begin{tabular}{lcccccccccccccc}
		\hline
		 $\alpha_{2000}$& $\delta_{2000}$ & Hip & HD & P & $\sigma P$ & a & $\sigma a$ & $\pi_{1997}$ &$\sigma \pi$ & $\pi_{2007}$ & $\sigma \pi$ & $\pi_{2018}$ & $\sigma \pi$ \\
				\hline
  \space &\space &\space & \space & year& - & arc-seconds & -&  mas & - & mas& - & mas&- \\
      \hline				
		000000.91& -192955.8 & 2&	224690 & 1.369	& 0.052& 0.0143&	0.00281&	21.9&	3.1&	20.85&	1.13&	25.121&	0.32 \\
		000019.1&	-441726&	25&	224750&	384.1&	22.5&	1.023&	0.096&	13.74&	0.98&	12.29&	0.77&	8.15&	0.665\\
		000034.35&	-530551.8&	50&	224782&		948.6&	284.6&	2.17&	0.43&	16.89&	0.8&	16.83&	0.51& 16.35&	0.036\\
		000123.67&	393638.2&	110&	224873&	223.2&	12.2&	0.8798&	0.0039&	20.42&	1.91&	20.15&	0.89&	19.27&	0.07\\
		000208.72&	-681650.6&	169&	224953&	290&	&	2.738&	&	63.06&	1.98&	65.24&	1.76&	58.96&	0.028\\
		000210.18&	270455.6&	171&	224930&	26.28&  & 	0.83&	&	80.63&	3.03&	82.17&	2.23&	79.07&	0.56\\
		000225.33&	104635.9&	190&	224994&	129.72&	&	0.366&	&	11.74&	0.53&	11.43&	0.93&	4.899&	0.75\\
		000238.92&	-82916.6&	210&	225015&	414.95&	20.96&	0.505&	0.021&	7.65&	1.93&	7.42&	1.45&	6.46&	0.62\\
		000247.14&	20749.4&	223&	225028&	1117&	&	3.14&	&	21.58&	1.65&	23.46&	0.91&	23.2&	0.05\\
		000446.86&	381025.2&	385& \space&	403&	&	0.74&	&	9.37&	2.81&	10.81&	2.14&	11.26&	0.08\\
		000529.06&	340620.4&	461&	39&	 47.06&	&	0.138&	&	11.04&	0.91&	10.3&	0.75&	8.36&	0.41\\
		000541.03&	454843.3&	473&	38&	1550.637&	&	11.7613&	 &	85.1 &	2.74 &	88.44 &	1.56&	86.87&	0.05\\
		000541.03 &	454843.3&	473&	38&	509.65&	96.99&	6.21&	0.77&	85.1&	2.74&	88.44&	1.56&	86.87&	0.05\\
		000541.03&	454843.3&	473&	38&	83037&	&	205.1&	&	85.1&	2.74&	88.44&	1.56&	86.87&	0.05\\
		000615.81&	582612.5&	518&	123&	106.7&	&	1.44&	&	49.3&	1.05&	46.56&	0.65&	47.80&	0.044\\
		000828.39&	345604.3&	689&	375&	12.81&	&	0.09&	&	12.72&	0.86&	11.69&	0.67&	10.11&	0.46\\
		000850.82&	864716.3&	705&	245&	0.19603&	&	0.0025&	7E-4&	16.42&	0.7&	17.51&	0.69&	18.03&	0.065\\
		000915.63&	251656.3&	754&	471& 1.323&	0.035&	0.0098&	0.0028&	22.07&	2.31&	19.45&	1.4&	24.89&	0.39\\
		000920.18&	794252.4&	760&	431&	540&	&	0.995&	&	8.56&	1.17&	9.48&	0.54&	9.32&	0.08\\
		000921.02&	-275916.5&	761&	493&	616.04&	&	1.5&	&	14.57&	1.34&	12.91&	0.72&	14.51&	0.15\\
		001005.26& 382453.6&	823&	&25.912&	0.232& 0.1121& 0.0017&	10.83&	2.1& 13.39&	2.07&	11.30&	0.08\\
		001038.56& -731327.7&	865&	661& 1637.7& 84& 2.762&	0.177&	15.06& 0.7&	14.2& 0.46&	18.34&	1.16\\
		001210.31& 464629.7&	984&	764&	269.52&	&	0.26& &	4.88& 1.22&	3.74& 1.11&	4.54& 0.20\\
		001230.12& 143349.3& 999&		&3.428& 0.242& 0.01839&	0.00124& 24.69&	1.2& 24.38&	0.95& 22.68& 0.201\\
		001323.93& 265915.4& 1076&	895& 421.98& 7.92&	0.641&	0.003&	8.08&	1.15&	7.61&	0.97&	18.69&	0.753\\
			\hline
	\end{tabular}
\end{sidewaystable*}

We investigated how measurements of the dynamical masses of the selected CVBSs are affected by the distances inferred from the parallax measurements in each of the catalogues discussed above.  We also compared these results with masses estimated from two different indirect methods; Malkov's photometric masses \citep{2012yCat..35460069M, 2012A&A...546A..69M} and  Al-Wardat's multi-parameter approach for analyzing CVBSs ~\citep{2002BSAO...53...51A, 2002BSAO...53...58A, 2007AN....328...63A, 2014AstBu..69...58A,2014PASA...31....5A,2016RAA....16..166A,2017AstBu..72...24A,2018arXiv180203804M,2018JApA...39...58M}. The latter is a computationally complex method employing colours, colour indices and magnitude differences of the system along with its parallax to build individual synthetic spectral energy distributions (SED) for each component of the system. From this, a complete set of physical and geometrical parameters can be deduced for each star. The method makes use of  Kurucz (ATLAS9) line-blanketed plane-parallel model atmospheres of the individual components ~\citep{1994KurCD..19.....K}.

Comparing dynamical with photometric masses is of specific importance in estimating empirical astrophysical equations and judging the accuracy of the zero points and constants. It also gives important information about the accuracy of orbital parameters of binary stars (BS) and multiplicity ratio among all-stars \citep{duchene2013stellar}.
Once the physical and geometrical parameters of the stars have been estimated, especially log $\rm L $ and log $ T_{\rm eff}$,  the positions of the individual components of the system can be located on the Hertzsprung-Russell (H-R) diagram. Hence, their masses can be estimated using evolutionary tracks such as those of \cite{2000yCat..41410371G}.

\section{Hipparcos and Gaia Observations of visual binary parallaxes}
  Parallaxes of both multiple stellar systems and even single stars are important in providing distance estimates that help in determining the stellar physical and geometrical parameters, especially the masses.
  Ideally, parallaxes should be measured geometrically, a model-independent method. However, for many stars they have only been available indirectly from photometric or spectroscopic observations and these are dependent on stellar atmosphere modelling.
  The advent of space-based astrometry has increased the sample size and accuracy of geometric parallax measurements, which presents an important opportunity to compare these with those determined by other methods.
  Furthermore, we can examine how better parallax measurements can reduce the errors in stellar mass determinations, leading to improved tests of the mass-luminosity and mass-radius relations and related stellar formation and evolution theories.

\begin{figure}
\minipage {0.42\textwidth}
  \includegraphics[width=\linewidth]{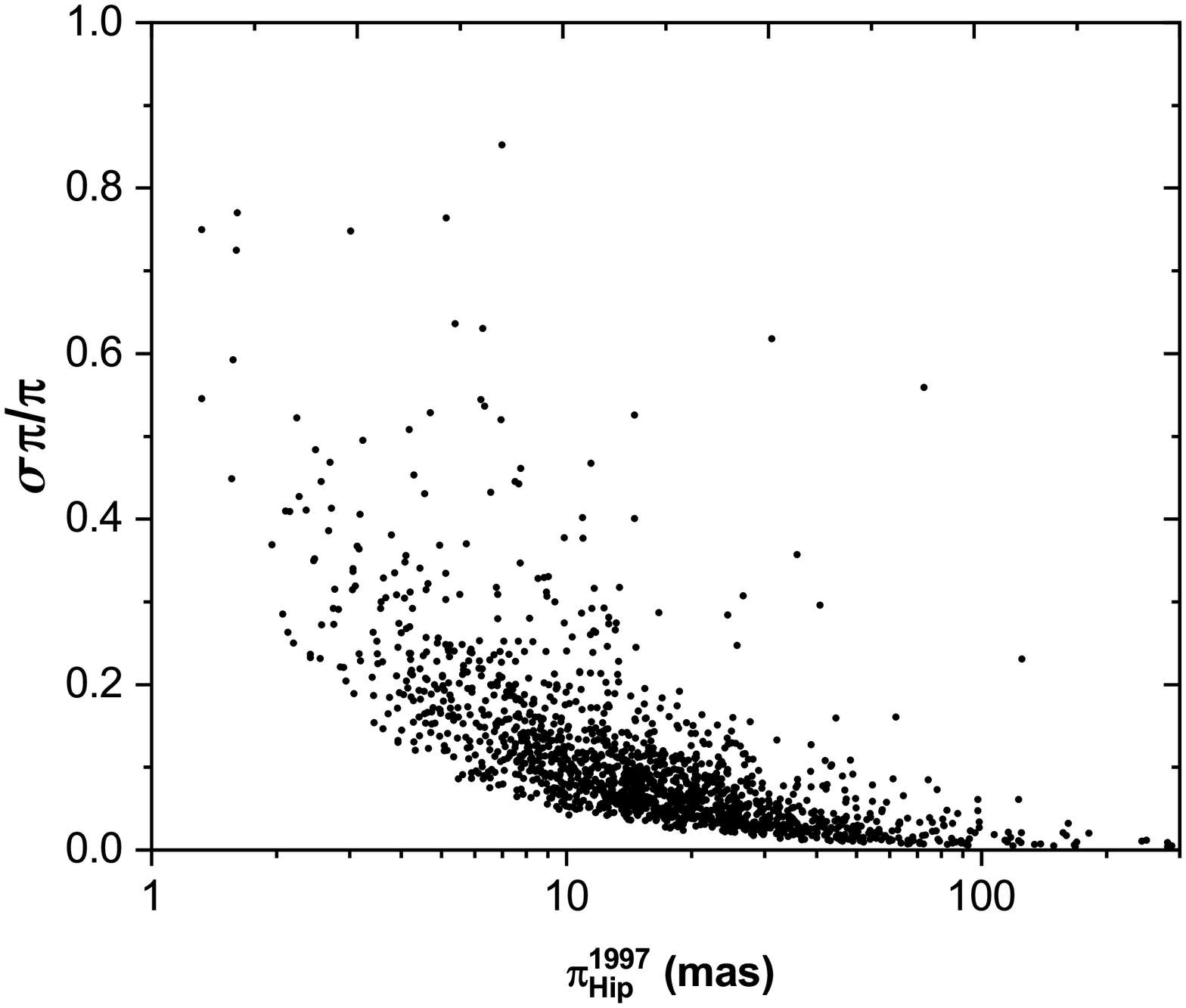}
  \caption{ The fractional errors for Hipparcos 1997 vs parallax measurements  distribution .}\label{log97}
\endminipage\hfill
\minipage{0.42\textwidth}
  \includegraphics[width=\linewidth]{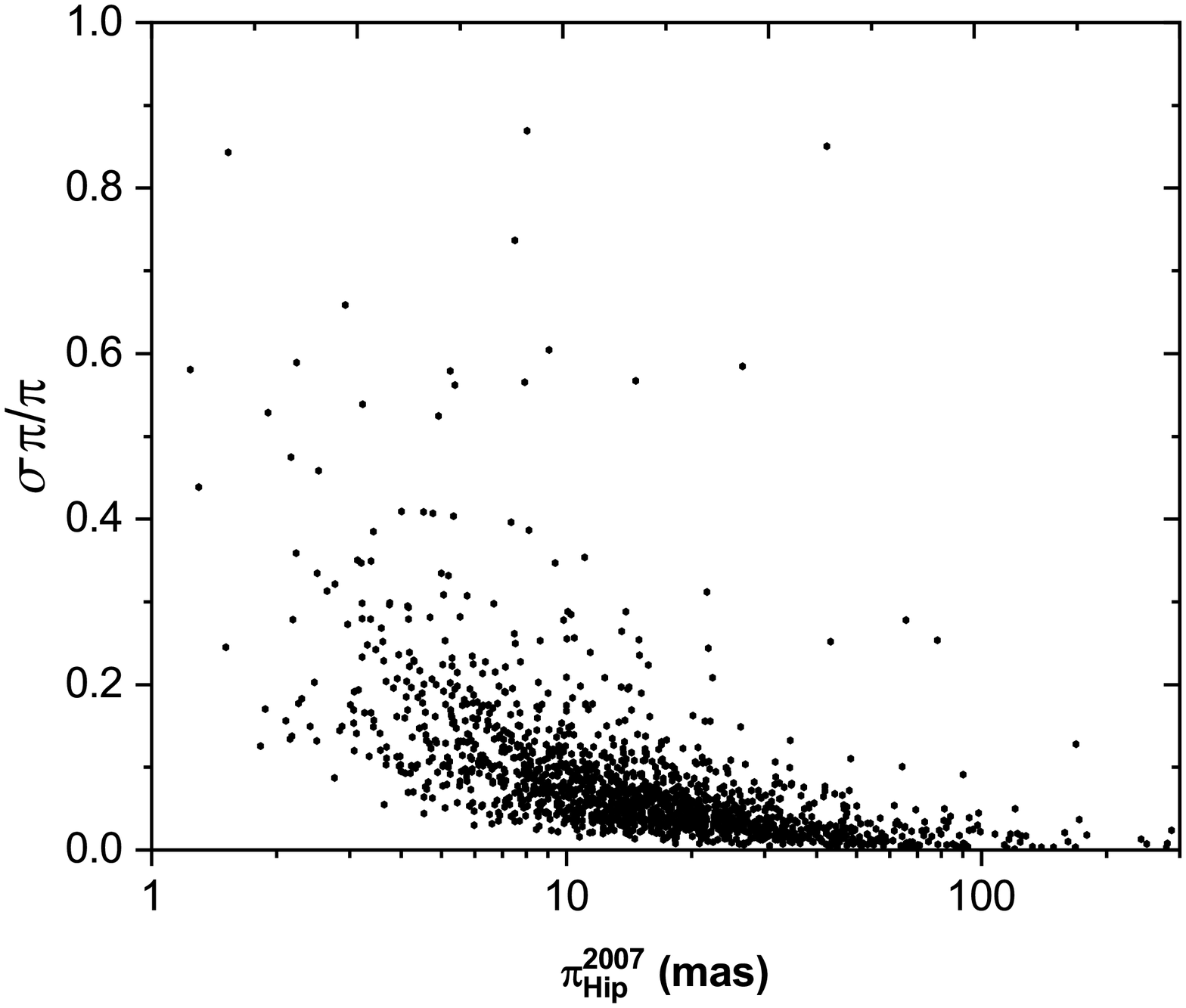}
  \caption{The fractional errors for Hipparcos 2007 vs parallax measurements  distribution.}\label{log07}
\endminipage\hfill
\minipage{0.42\textwidth}%
  \includegraphics[width=\linewidth]{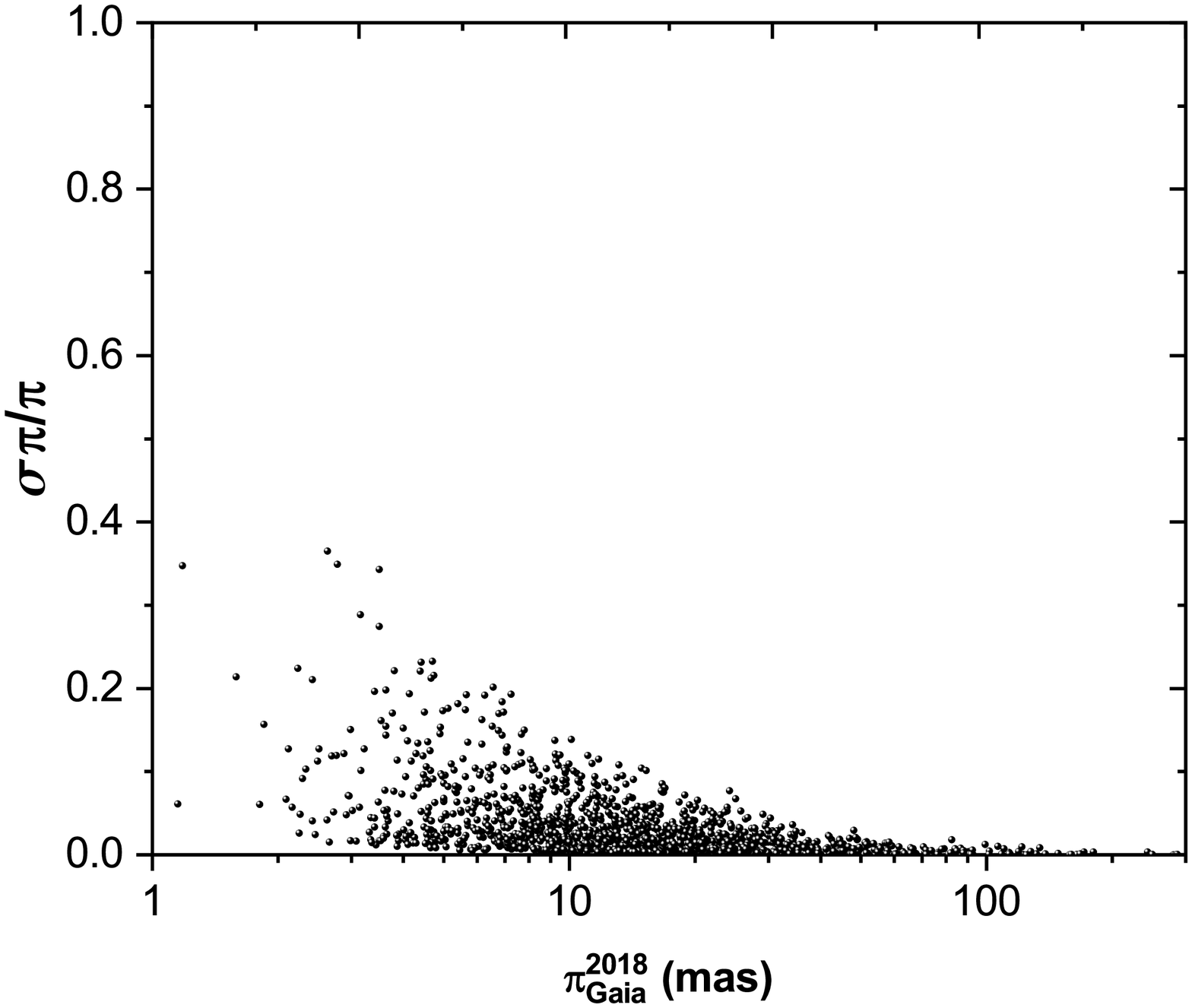}
  \caption{The fractional errors for Gaia DR2 2018 vs parallax measurements  distribution.}\label{log18}
\endminipage
\end{figure}

\begin{figure}
	\includegraphics[width=8cm]{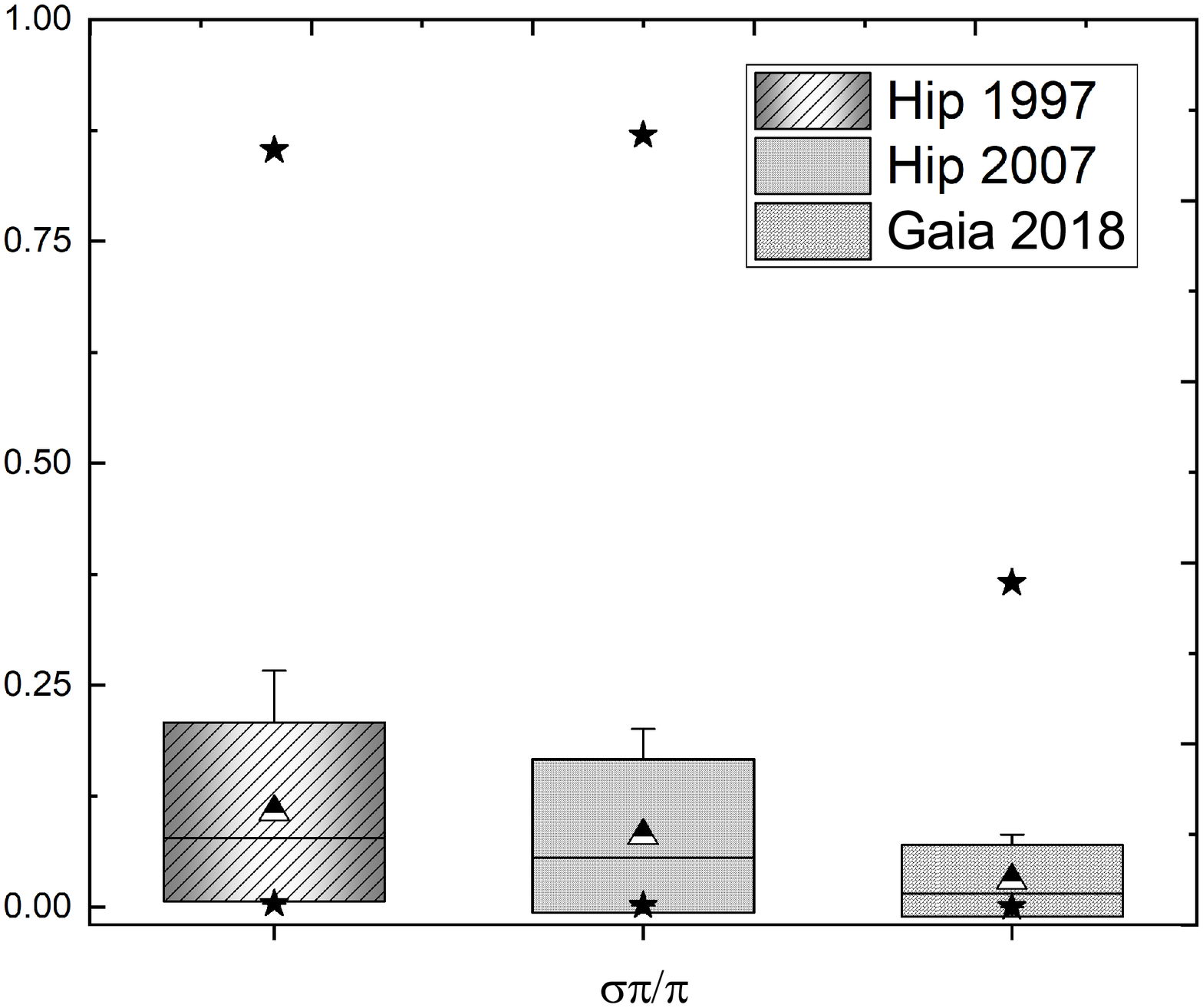}
    \caption{Statistical box chart diagram of the fractional errors of parallax measurements of the three catalogues; Hip 1997, Hip 2007, and Gaia DR2 2018. This display the distribution of data based on a five-number summary (“minimum,”  median, Mean, SD and “maximum”).}
    \label{pllx ana}
\end{figure}

\subsection{van Leeuwen 2007 reanalysis of the Hipparcos data}

In 1999, two years after the release of Hipparcos and Tycho Catalogues, \cite{narayanan1999correlated} studied the correlation of Hipparcos parallax measurements for  Pleiades and  Hyades clusters. They noted that the parallaxes were larger on average than the other reported values for the stars of these two clusters. Later, Makarov  showed an inconsistency between the mean parallax of the Pleiades cluster from Hipparcos catalog and that obtained from stellar evolution theory and photometric measurements \citep{2002AJ....124.3299M}.
In 2005, measurements made using the Hubble Space Telescope (HST) fine guidance sensor (FGS) confirmed the error in the Hipparcos parallax of the Pleiades \citep{2005AJ....129.1616S}.

A new reduction of the raw Hipparcos data depending on dynamical modelling of the satellite's attitude was developed by van Leeuwen and Fantino \citep{2005A&A...439..791V}, with a full reanalysis of the Hipparcos data published in 2007 \citep{2007AA...474..653V}. The latter paper claimed a parallax accuracy up to a factor of 4 better than the original catalogue for nearly all stars brighter than 8 magnitude. Nevertheless, the revised Hipparcos measurement of the distance to the Pleiades of $120.0 \pm 1.9$ pc \citep{2009A&A...497..209V, 2019MNRAS.487.3568S} remains anomalous compared to
estimates based on isochrone fits of the stellar photometry \citep{1993A&AS...98..477M, 2001A&A...374..105S} and results obtained using eclipsing binaries by \citep{2004A&A...425L..45Z, 2005A&A...429..645S}, which placed their distances in the range $(130-137)$ pc. These are in good agreement with the $134.6\pm0.6$ pc distance obtained from Gaia DR2 \citep{2018yCat.1345....0G}.

\subsection{The visual Binary sample and parallax data}

The binary systems selected for this study (Table~\ref{orb}) have to fulfil two main requirements: they should have parallax measurements in all three space-based astrometric catalogues and a solved orbit in the ORB6 catalogue.
Systems with zero or negative parallaxes were excluded.

Table~\ref{orb} shows the first 25 lines of the sample - the complete table with 1710 stars is available in electronic format.  The first four columns give information about the star;  Right Ascension $\alpha_{2000}$,  Declination $\delta_{2000}$, Hipparcos and  HD names. Columns 5,6,7 and 8 give the orbital period $P$, error of the orbital period $\sigma P$ , the semi-major axis $a$, and error of semi-major axis $\sigma a$, all as given in the ORB6 catalog. The last six columns list the trigonometric parallax of the stars with their errors as given by Hipparcos 1997 ($\pi_{1997}$) ~\cite{1997yCat.1239....0E}, van Leeuwen reduction ($\pi_{2007}$)  ~\citep{2007AA...474..653V}, and Gaia DR2 ($\pi_{2018}$) ~\cite{2018yCat.1345....0G}.

The distribution of parallax measurements for each catalogue, as a function of the fractional parallax error ($\dfrac{\sigma \pi}{\pi}$), are shown in Fig.~\ref{log97} (Hip 1997), Fig.~\ref{log07} (Hip 2007), and Fig.~\ref{log18} (Gaia DR2 2018).
The much better measurement errors in Gaia DR2 are  evident, but an improvement in the Hipparcos 2007 data, compared to Hipparcos 1997, can also be seen, particularly for stars with low parallax.
This is further illustrated in Fig.~\ref{pllx ana} which shows the fractional parallax measurement errors of the sample for each catalogue, expressed as a (median $\pm$ standard deviation) in a box chart diagram; stars- Minimum and Maximum values; Triangle- Mean values.

\begin{figure}
	\includegraphics[width=7cm]{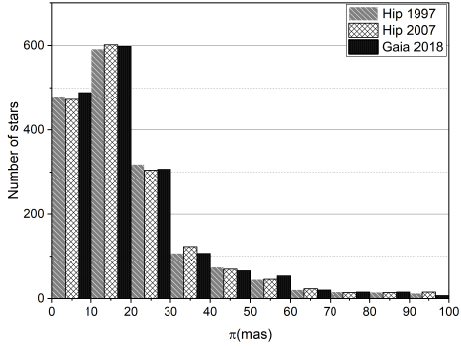}
    \caption{Distribution of the number of measured binary systems within the range ($0<\pi\le 100$ mas) of the three catalogues; Hip 1997(gray bars), Hip 2007 (white bars), and Gaia DR2 2018 measurements (black bars).}
    \label{114}
\end{figure}

\begin{figure}
	\includegraphics[width=7cm]{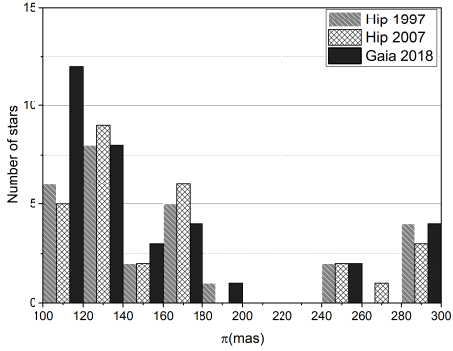}
    \caption{Distribution of the number of measured binary systems within the range ($100 \le \pi\le 300$ mas) of the three catalogues;  Hip 1997(gray bars), Hip 2007 (white bars), and Gaia DR2 2018 measurements (black bars).}
    \label{146}
\end{figure}

A further consideration in comparing the parallax measurements is whether or not the parallax values are generally in agreement with each other (within the errors).
The distribution of the number of measured binary systems within specific parallax bins is a way of illustrating this. Fig.~\ref{114} shows the number of stars within 10 mas bins for the parallax range $\pi$ $\sim$ ($0<\pi\le 100$ mas).
There is consistency between the catalogues with small deviations, likely representing the few stars that are distributed into adjacent bins where there are small changes in value close to the bin boundaries.
While there is no attempt to select a statistically complete sample in this work, the figure clearly shows that the distribution of systems reflects the accessible volume, with majority of the binary systems lying farther than 30 pc ($\pi\le 30$ mas), with a peak between 10 and 20 mas.
There is a decline in the number of binary systems below 10 mas, probably linked to the declining ability to resolve binaries at increasing distance.
Fig.~\ref{146} shows the distribution for nearby systems within 20 mas bins and parallax measurements on the range $\pi$ $\sim$ ($100\le \pi\le 300$) mas. This shows some differences in parallax measurements, but the counting statistical errors are similar in size to the differences.

\begin{figure}
	\includegraphics[width=7cm]{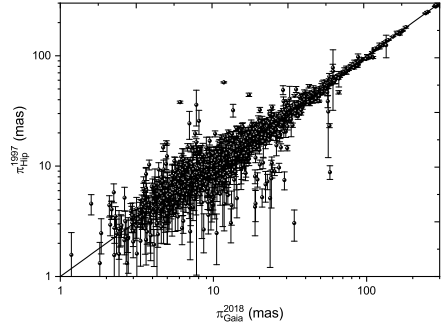}
    \caption{ Scatter plots of parallax measurements of  Hip 1997 parallaxes with errors vs. Gaia (DR2) 2018 parallax measurements. The line represent y=x.}
    \label{err197}
\end{figure}

\begin{figure}
	\includegraphics[width=7cm]{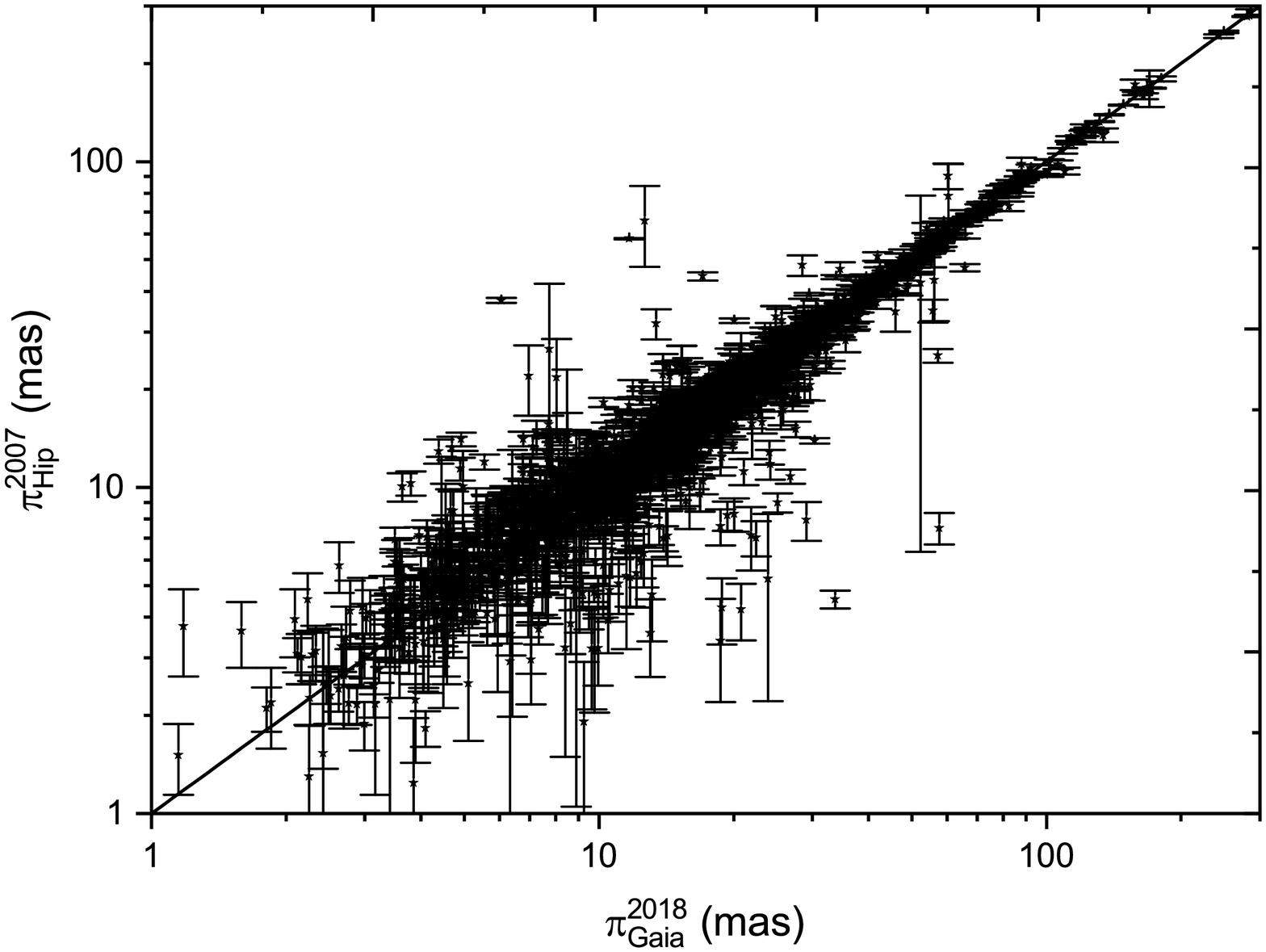}
    \caption{Scatter plots of parallax measurements of  Hip 2007 parallaxes with errors vs.Gaia (DR2) 2018 parallax measurements. The line represent y=x.}
    \label{err07}
\end{figure}

Fig.~\ref{err197} and Fig.~\ref{err07} show scatter plots (with errors marked) comparing the two catalogues of Hipparcos trigonometric parallax measurements with the Gaia DR2 data. Both plots show the y=x line of equal parallax.
There is good agreement between Gaia DR2 and both treatments of the Hipparcos data. However, there are several stars where there are significant differences between Gaia and Hipparcos measurements. These are the same objects for both Hipparcos treatments.

\begin{figure}
\minipage {0.42\textwidth}
  \includegraphics[width=\linewidth]{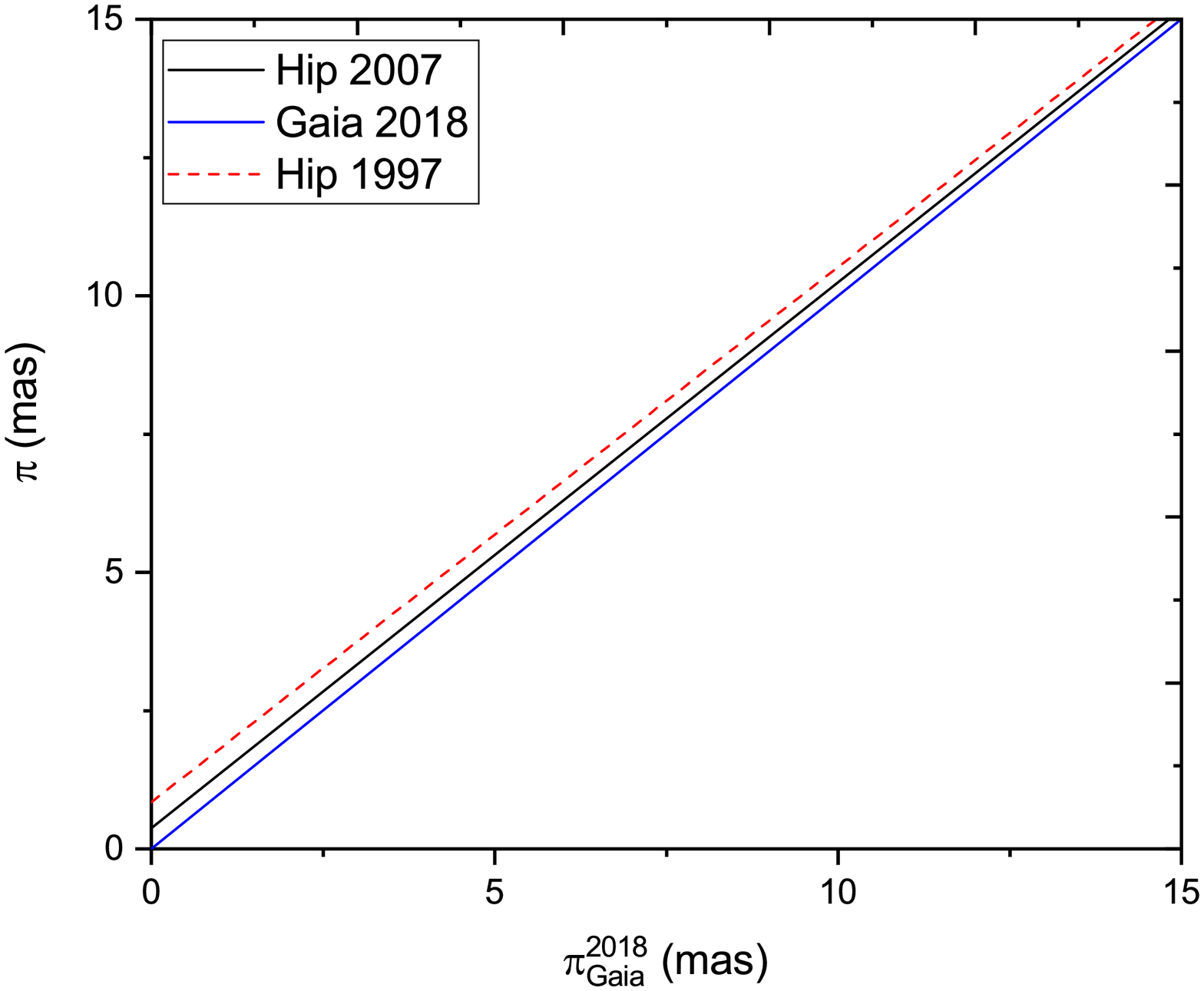}
  \caption{Trigonometric parallax of Gaia 2018 vs that of Hipparcos 1997 and Hipparcos 2007(Van Leeuwen reduction) for ($0 \le \pi\le 15$) mas.}\label{355}
\endminipage\hfill
\minipage{0.42\textwidth}
  \includegraphics[width=\linewidth]{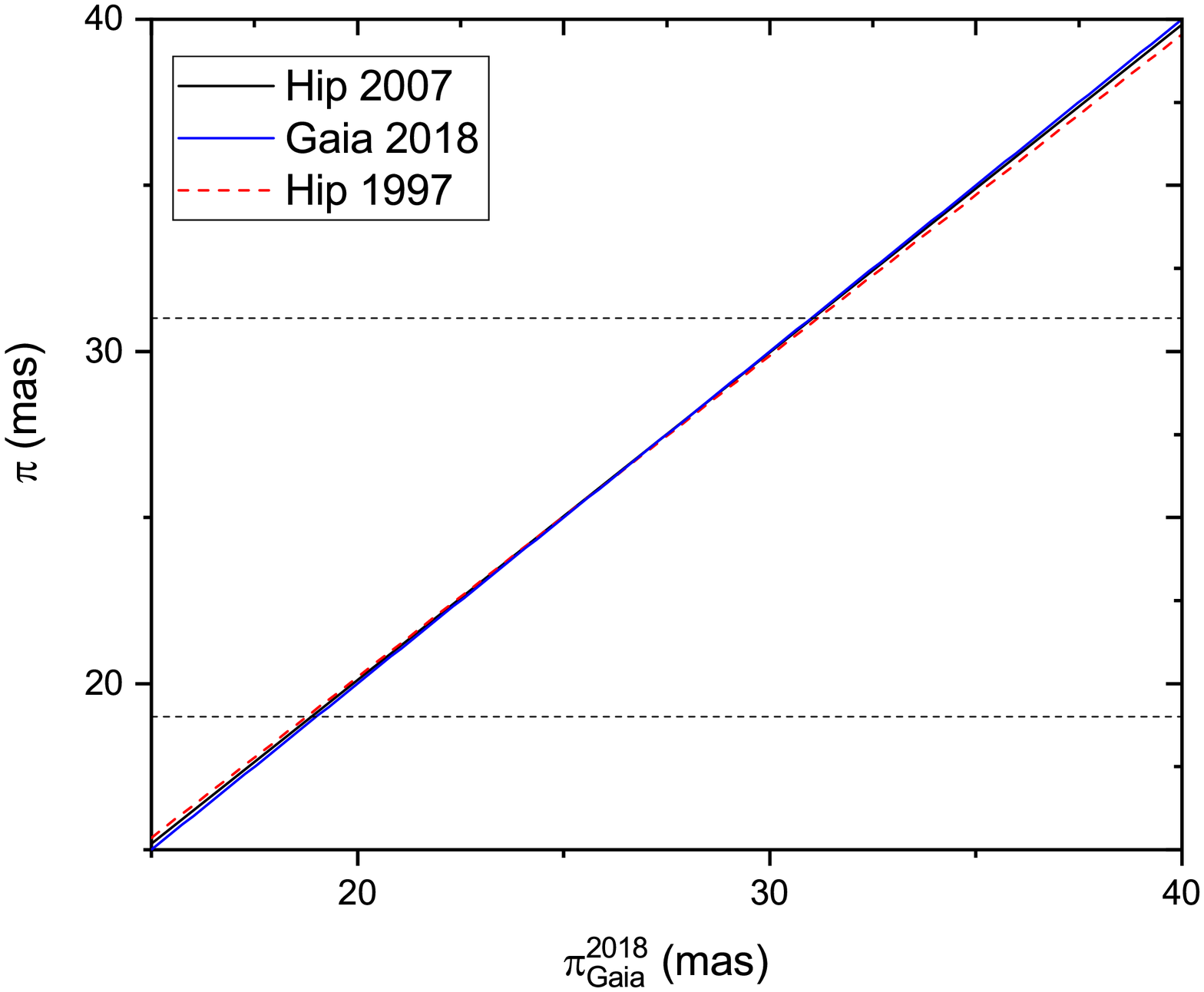}
  \caption{Trigonometric parallax of Gaia 2018 vs that of Hipparcos 1997 and Hipparcos 2007(Van Leeuwen reduction) for ($15 \le \pi\le 40$) mas.}\label{3551}
\endminipage\hfill
\minipage{0.42\textwidth}%
  \includegraphics[width=\linewidth]{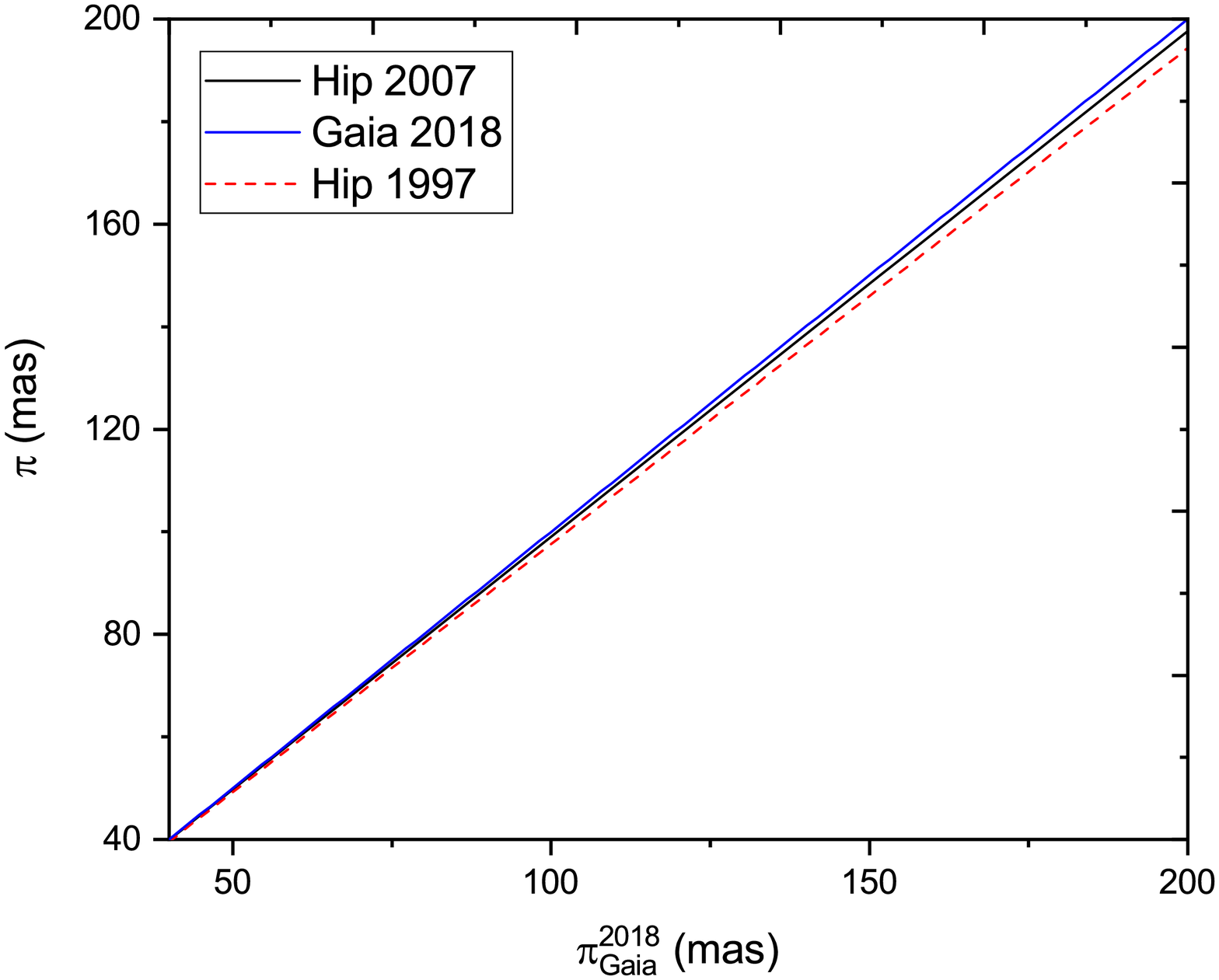}
  \caption{Trigonometric parallax of Gaia 2018 vs that of Hipparcos 1997 and Hipparcos 2007(Van Leeuwen reduction) for ($40 \le \pi\le 200$) mas.}\label{3552}
\endminipage
\end{figure}

\begin{figure}
	\includegraphics[width=7cm]{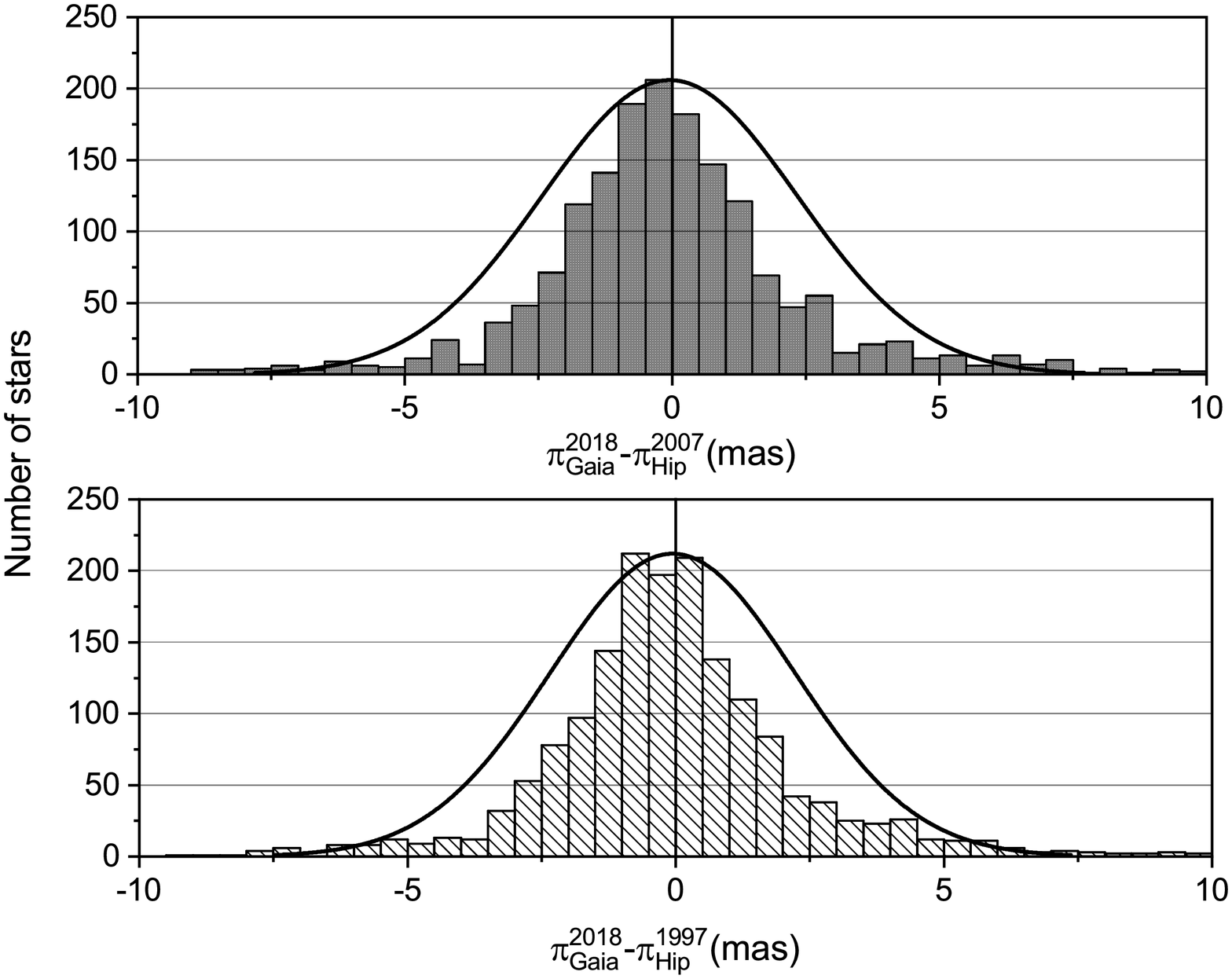}
    \caption{Distribution of $\Delta \pi$, (top) is ($\pi_{Gaia}^{2018}-\pi_{Hip}^{2007}$), the (bottom) is ($\pi_{Gaia}^{2018}-\pi_{Hip}^{1997}$), the line shows a Gaussian distribution calculated for the observed peaks.}
    \label{gauus}
\end{figure}

To search further for any trends in the parallax comparison, we plotted the best fit straight line to the parallax measurements of Hip 1997 and Hip 2007 against Gaia DR2 for several different parallax intervals.
Fig.~\ref{355} shows the parallax range 0-15 mas, Fig.~\ref{3551} shows the parallax range  15-40 mas, and Fig.~\ref{3552} shows the parallax range  40-200 mas. The y=x lines of equal parallax are in blue.
The best agreement between Hipparcos and Gaia is found in the 10-40 mas range. For lower parallaxes (greater distances) the Hipparcos measurements are systematically higher than Gaia, while the situation is reversed for the higher parallaxes. The weighted mean offset in the whole sample of parallax measurements (Fig.~\ref{gauus}) of ($\pi_{Gaia}^{2018}-\pi_{Hip}^{2007}$) is $-43.32 \mu as$, while for ($\pi_{Gaia}^{2018}-\pi_{Hip}^{1997}$) it is $-59.03 \mu as$.
The analysis shows clearly that the Van Leeuwen 2007 analysis is in better agreement with Gaia DR2 than the original Hipparcos reduction, justifying the reworking of the astrometric data. However, it is also clear that this work has now been superceded by Gaia DR2 in terms of astrometric accuracy.

One important factor to note in the Gaia DR2 data is that all objects in the catalogue are effectively single stars. Single line spectroscopic binaries were treated as single objects while double line binaries detected as such were excluded from the catalogue \citep{2018A&A...616A...1G}. For resolved systems, such as those included in this analysis, binary motion could affect the accuracy of the parallax determination for orbital periods below 2 years \citep{2018A&A...616A...1G}. In their analysis of the stellar parameters using Gaia DR2 data, \cite{2018A&A...616A...8A} noted some of these issues and recommended only using estimates of radii and luminosities for stars with fractional parallax uncertainties of 20\% or less.

We found that some BSs have  parallax discrepancies between Gaia and either Hipparcos catalogues. A number have significant differences  (larger than 5 mas) in parallax measurements between the two catalogues: Hip (190, 1076, 1349, 1392, 1625, 4849, 7968, 12301, 19365, 19832, 20605, 20777, 20807, 33645, 41423, 43422, 45858, 48940, 58001, 58057, 58669, 59776, 59780, 59816, 60091, 60665, 62179, 62124, 64175, 65026, 66077, 68148, 68170,  72896,  77098, 79787, 80117, 80816, 84092, 84976, 86400, 87044, 88964, 90851, 96280, 99965, 103019, 104256, 106255, 106985, 110088, 111685, 113697, 118209, 118266).
The study of these systems is important and can be used as a tool to judge the reliability of Gaia parallax measurements. As an example, we analyzed the system Hip 68170 using Al-Wardat's method (Section~\ref{specific systems}).

The reasons for the large differences in parallax measurements between Hipparcos and Gaia could be due to either; the effect of the interstellar extinction, or the effect of the change of photo-center of these binaries as noted by several authors~\citep{1998AstL...24..673S, 2020AcA....70...19L}.

\section{Masses}
\subsection{Dynamical Masses}
Binary stars are the best source of information regarding stellar masses, where we can calculate the dynamical mass sum of a BSs once we have its orbit and parallax \cite{2014MNRAS.444.3641D}.
DR2 gives more accurate parallax measurements, which in principle means more accurate masses and fundamental parameters.
There is a clear strong relation between the multiplicity in stars in general,  and the estimated or calculated mass-sums. \cite{duchene2013stellar}  discussed the mass-dependence of the main sequence stars multiplicity properties and showed that the multiplicity rises significantly toward high-mass stars.

There are many methods  for solving and quantifying orbits of CVBSs, these methods include Kowalsky's method ~\citep{smart1930derivation}, Monet's method which applies Fourier transformation ~\citep{1979ApJ...234..275M}, Docobo's Analytical method \citep{1985CeMec..36..143D}, and Tokovinin's dynamical method \citep{1992ASPC...32..573T}.

In this study, we want to evaluate the impact of improved parallax data on the mass values and accuracy of the measurements of CVBSs. Therefore, we recalculated the dynamical masses of all CVBSs with solved orbits in the ORB6 using parallax measurements of the three astrometric catalogues discussed in section 2 and the photometric mass sums from \citep{2012A&A...546A..69M}.
%and Kepler's third law from equations~(\ref{dynamical mass} and \ref{E dynamical mass}).
Results are listed in Table~\ref{mass}.

\begin{equation}
      \mathcal{M}_{dyn}=\left(\frac{a}{\pi}\right)^3\frac{\mathcal{M}_\odot}{P^2}.
	\label{eq:quadratic}
\end{equation}

where $P$ is the orbital period (in years), $\mathcal{M}_{dyn}$ is the dynamical mass sum in
solar mass $\mathcal{M}_\odot$, $a$ and $\pi$ are the semi-major axis and the parallax
in arcsec, respectively.

\begin{equation}
\frac{\sigma_{\mathcal{M}_{dyn} }} {\mathcal{M}_{dyn}}= \sqrt{9\left(\frac{\sigma_\pi}{\pi}\right)^2 +9\left(\frac{\sigma_a}{a}\right)^2 +4\left(\frac{\sigma_P}{P}\right)^2}\
	\label{E dynamical mass}
\end{equation}

\begin{sidewaystable*}
	\centering
	\caption{Dynamical mass sum with their uncertainties and Malkov photometric mass sum. The first twenty five lines of the table.}
	\label{mass}
	\begin{tabular}{lccccccccccccr}
		\hline
		Hip& $\pi_{1997}$ &$\sigma \pi$ & $\mathcal{M}_{dyn}^{1997}$  & $\sigma \mathcal{M}$& $\pi_{2007}$ & $\sigma \pi$ & $ \mathcal{M}_{dyn}^{2007}$  & $\sigma \mathcal{M}$& $\pi_{2018}$ & $\sigma \pi$ & $\mathcal{M}_{dyn}^{2018}$  & $\sigma \mathcal{M}$ & $\mathcal{M}_{ph}$ \\
		
		\space& mas & - & $\mathcal{M}_\odot$& - &mas & - & $\mathcal{M}_\odot$& - & mas & - &$\mathcal{M}_\odot$&-&$\mathcal{M}_\odot$\\
		\hline
	2&	21.9&	3.1&	0.149&	0.109	&20.85&	1.13&	0.172&	0.106&	25.121&	0.319&	0.099&	0.059&\\
	25&	13.74&	0.98&	2.797&	1.042&	12.29&	0.77&	3.909&	1.40&	8.144&	0.665&	13.434&	5.255&\\
	50&	16.89&	0.8&	2.357&	2.018&	16.83&	0.51&	2.38&	2.024&	16.351&	0.036&	2.598&	2.194&\\
	110&	20.42&	1.91&	1.605&	0.484&	20.15&	0.89&	1.671&	0.288&	19.269&	0.07&	1.911&	0.212& 1.79\\
	169&	63.03&	1.98&	0.975&	0.092&	65.24&	1.76&	0.879&	0.071&	58.962&	0.028&	1.191&	0.002&\\
	171&	80.63&	3.03&	1.579&	0.178&	82.17&	2.23&	1.492&	0.122&	79.069&	0.562&	1.675&	0.036& 1.58\\
	190&	10.47&	1.17&	2.539&	0.851&	11.43&	0.93&	1.951&	0.476&	4.899&	0.751&	24.779&	11.382&\\
	210&	7.65&	1.93&	1.671& 	1.293&	7.42&	1.45&	1.831&	1.113&	 6.459&	0.621&	2.776&	0.915&\\
	223& 21.58&	1.65&	2.469&	0.566&	23.46&	0.91&	1.922&	0.224&	23.199&	0.051&	1.987&	0.0132&\\
	385&	9.37&	2.81&	3.033&	2.729&	10.81&	2.14&	1.975&	1.173&	11.265&	0.076&	1.745&	0.035&\\
	461& 11.04&	0.91& 0.882& 0.218& 10.3& 0.75&	1.086&	0.237&	8.361&	0.409& 2.031& 0.298\\
	473&	85.1&	2.74&	1.098&	0.106&	88.44&	1.56&	0.978&	0.052&	86.874&	0.048&	1.032&	0.002&2.52\\
	473&	85.1&	2.74&	1.496&	0.809& 	88.44&	1.56&	1.333&	0.713&	86.874&	0.048&	1.406&	0.748&\\
	473&	85.1&	2.74&	2.03&	0.196&	88.44&	1.56&	1.809&	0.096&	86.873&	0.048&	1.908&	 0.003&\\
	518&	49.3&	1.05&	2.189&	0.139&	46.56&	0.65&	2.598&	0.109&	47.801&	0.044&	2.401&	0.007&\\
	689&	12.72&	0.86&	2.159&   0.438&   11.69&	0.67&	2.781&	0.478&	10.112&	0.457&	4.296&	0.582& 2.67\\
	705&	16.24&	0.7&	0.095&	0.081&	17.51&	0.69&	0.076&	0.064&	18.029&	0.065&	0.069&	0.058&\\
	754& 22.07&	2.31& 0.05& 0.046&	19.45&	1.4&	0.073&	0.065&	24.889&	0.394&	0.035& 0.030&\\
	760&	8.56&	1.17&	5.386&	2.208&	9.48&	0.54&	3.965&	0.678&	9.316&	0.082&	4.179&	0.111&3.22\\
	761 & 14.57&	1.34&	2.875&	0.793&	12.91&	0.72&	4.133&	0.691&	14.512&	0.146&	2.909&	0.088&\\
	823&	10.83&	2.1&	1.652&	0.964&	13.39&	2.07&	0.874&	0.408&	11.300&	0.076&	1.454&	0.077&\\
	865& 15.06&	0.7& 2.300& 0.595&	14.2& 0.46&	2.744& 0.655& 18.338&	1.159&	1.274&	0.368&\\
	984&	4.88&	1.22&	2.082&	1.562&	3.74&	1.11&	4.625&	4.118&	4.537&	0.197&	2.591&	0.337&\\
	999&	24.69&	1.2&	0.035&	0.010&	24.38&	0.95&	0.036&	0.009&	22.682&	0.201&	0.045&	0.011&\\
	1076&	8.08&	1.15&	2.804&	1.202&	7.61&	0.97&	3.356&	1.291&	18.69&	0.753&	0.226&	0.029&4.23\\	
	
		\hline
	\end{tabular}
\end{sidewaystable*}

\begin{figure}
	\includegraphics[width=8cm]{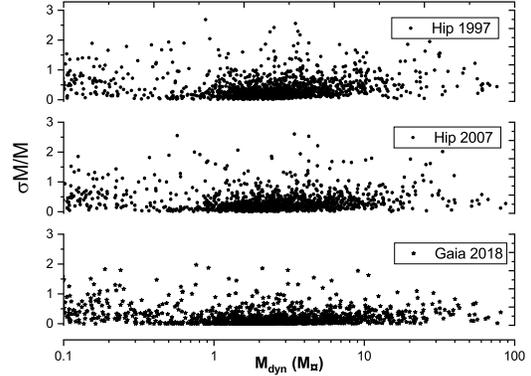}
    \caption{The distribution of dynamical masses which where calculated using Hip 1997, Hip 2007, and Gaia DR2 2018 parallax measurements against their fractional errors.}
    \label{mass distr}
\end{figure}

\begin{figure}

	\includegraphics[width=8cm]{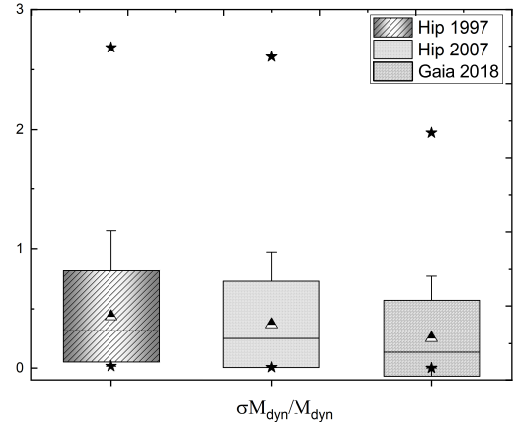}
    \caption{Statistical box chart analysis of the fractional errors of the dynamical mass sums which where calculated using Hip 1997, Hip 2007, and Gaia 2018 parallax measurements. This display the distribution of data based on a five-number summary (“minimum,”  median, Mean, SD and “maximum”).}
    \label{mass E distr}
\end{figure}

Fig.~\ref{mass distr} shows the distribution of dynamical masses  $\mathcal{M}$, which were calculated using Hip 1997, Hip 2007, and Gaia DR2 2018 parallax measurements, against their formal errors $\frac{\sigma \mathcal{M}}{\mathcal{M}}$. Note that systems with  high masses (greater than 100 $\mathcal{M}_{\odot}$, and $\frac{\sigma \mathcal{M}}{\mathcal{M}}$ less than 3) were excluded from the study. Fig.~\ref{mass E distr} shows the statistical box chart analysis of the formal errors of the dynamical mass sums expressed as a $Median \pm S.D$ in a box chart diagram; stars- Minimum and Maximum values; Triangle- Mean values.
The figures show that Gaia DR2 2018 has the lowest mean, and hence the higher accuracy, while Hip 1997 has the highest mean (i.e. the lowest accuracy), and Hip 2007 lies between them.

\begin{figure}
	\includegraphics[width=8cm]{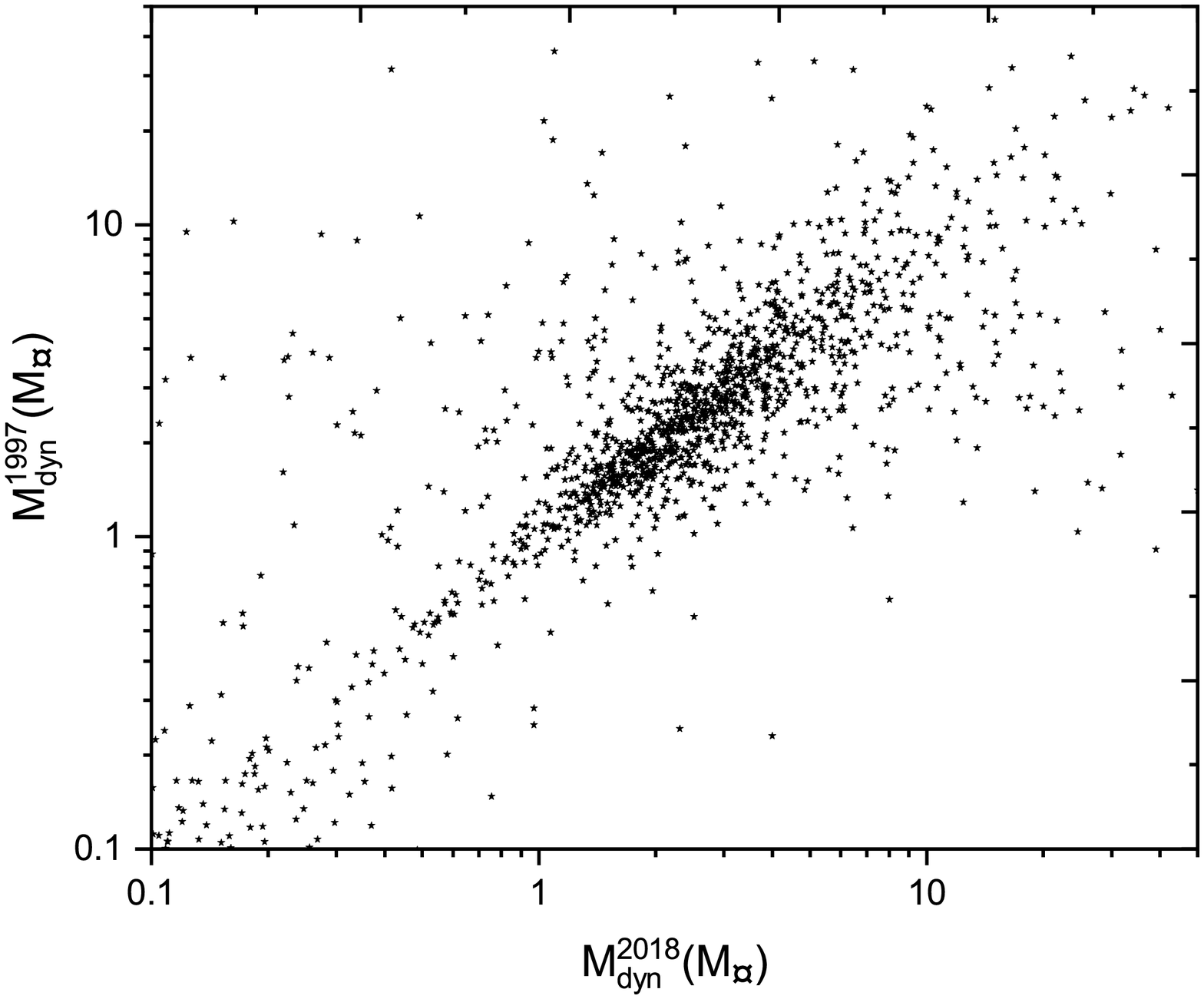}
    \caption{ Dynamical masses based on Hip 1997 parallax measurements vs. dynamical masses based on Gaia DR2 parallax measurements. }
    \label{madiff97}
\end{figure}

\begin{figure}
	\includegraphics[width=8cm]{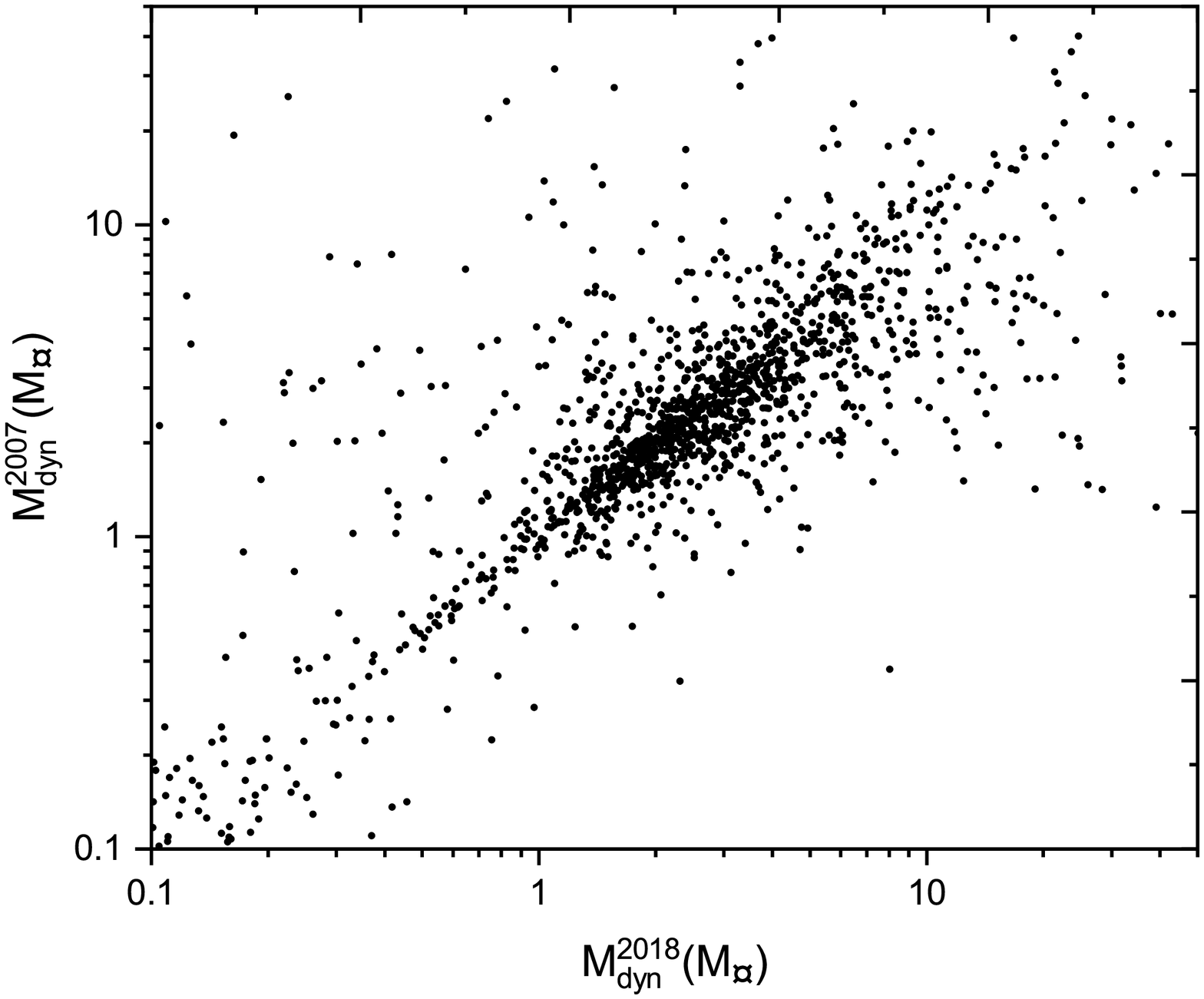}
    \caption{Dynamical masses based on Hip 2007 parallax measurements vs. dynamical masses based on Gaia DR2 parallax measurements.}
    \label{madiff07}
\end{figure}

Fig.~\ref{madiff97} and Fig.~\ref{madiff07} show scatter plots comparing the dynamical mass sum based on the two catalogues of Hipparcos trigonometric parallax measurements with the Gaia DR2 data.
These plots show a large scatter compared to the parallax plot from Fig.~\ref{err197} and Fig.~\ref{err07}, because any change in the parallax will be enlarged in the dynamical masses, based on Kepler's third law.

\subsection{\cite{2012A&A...546A..69M} Photometric Masses}
As previously stated, to get precise dynamical masses we need precise orbital parameters, which requires more relative positional measurements and accurate parallax measurements that are not always available. Therefore, it is important to have alternative validated methods for estimating stellar masses.

Malkov selected a sample of 652 visual binaries with good orbital solutions and used Hip 2007 parallax measurements \citep{2007AA...474..653V} to estimate luminosities and masses of the individual components of these BS.
He used the photometric empirical mass-luminosity $(M-L)$ relation, given as:
\begin{equation}
    m_{1,2}=f_{MLR}(m_{1,2}+5log \pi +5-A_v).
	\label{malkov eq}
\end{equation}
where $m_{1,2}$ are the apparent magnitudes of individual components, $f_{MLR}$ is the mass-luminosity relation, $A_v$ is the interstellar extinction value, and $\pi$ is the trigonometric parallax.

\begin{figure}
\minipage {0.42\textwidth}
  \includegraphics[width=\linewidth]{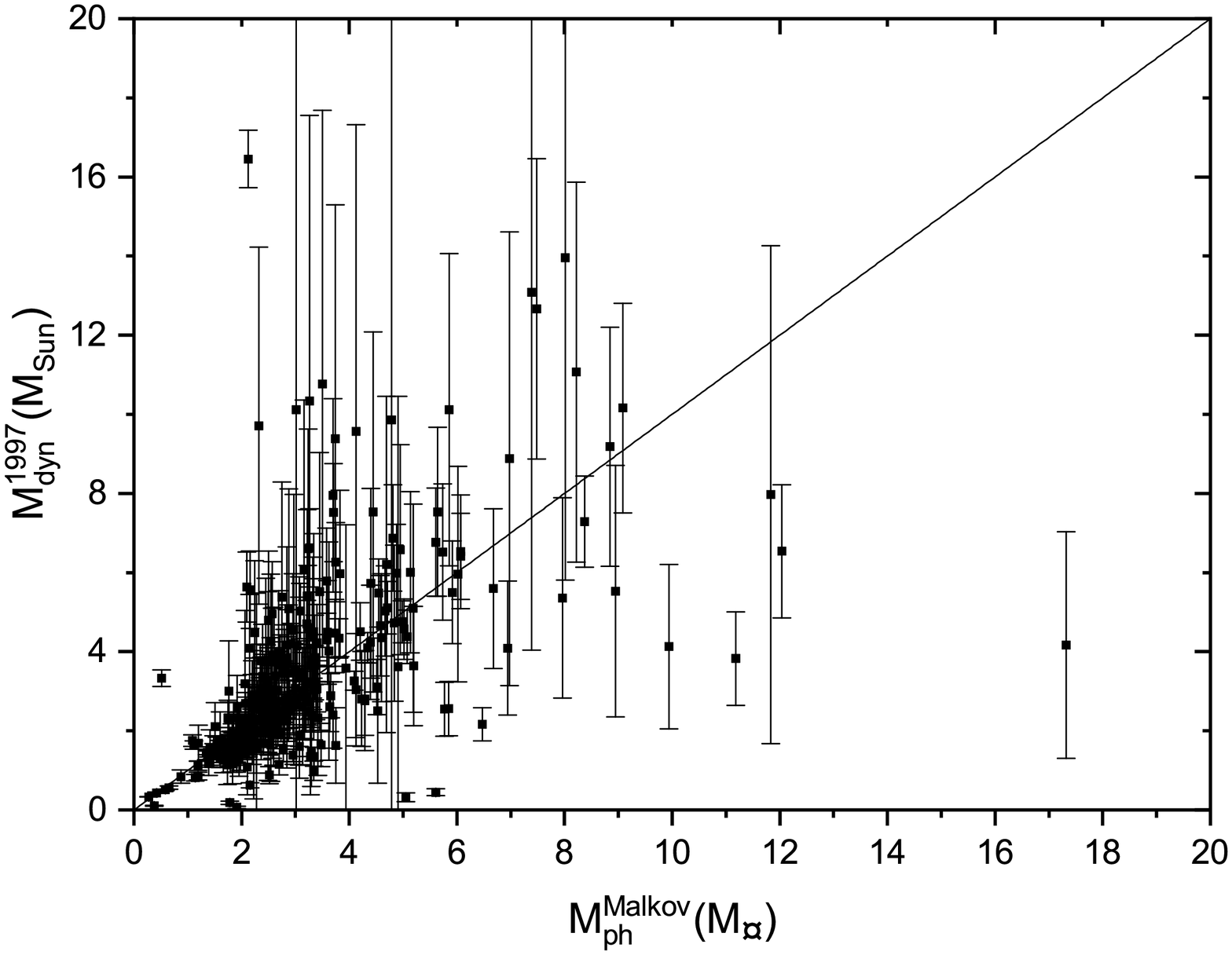}
  \caption{Dynamical mass depending on Hip 1997 parallax measurements vs. Malkov photometric mass.}\label{ehip 1997}
\endminipage\hfill
\minipage{0.42\textwidth}
  \includegraphics[width=\linewidth]{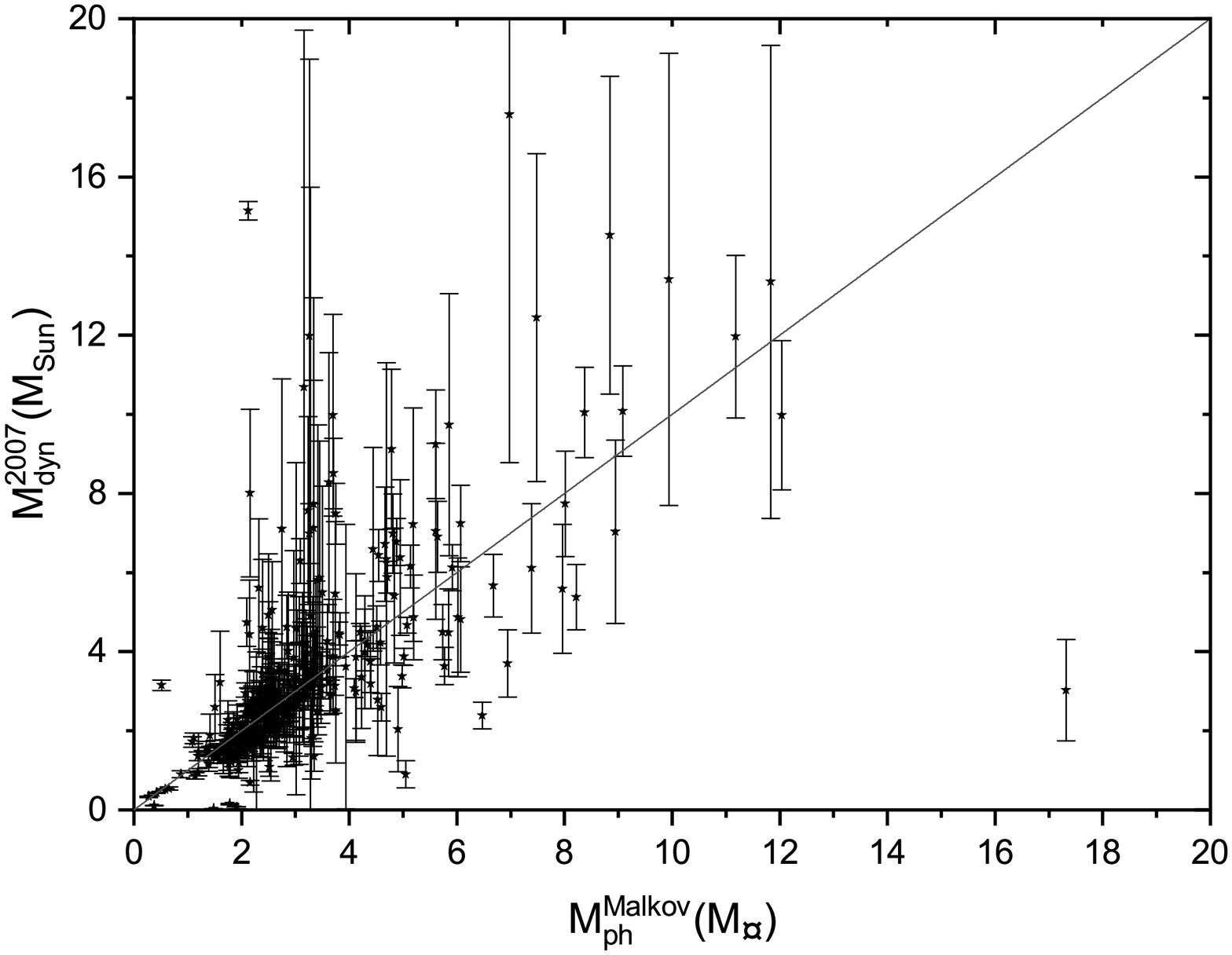}
  \caption{ Dynamical mass depending on Hip 2007 parallax measurements vs. Malkov photometric mass.} \label{ehip 2007}
\endminipage\hfill
\minipage{0.42\textwidth}%
  \includegraphics[width=\linewidth]{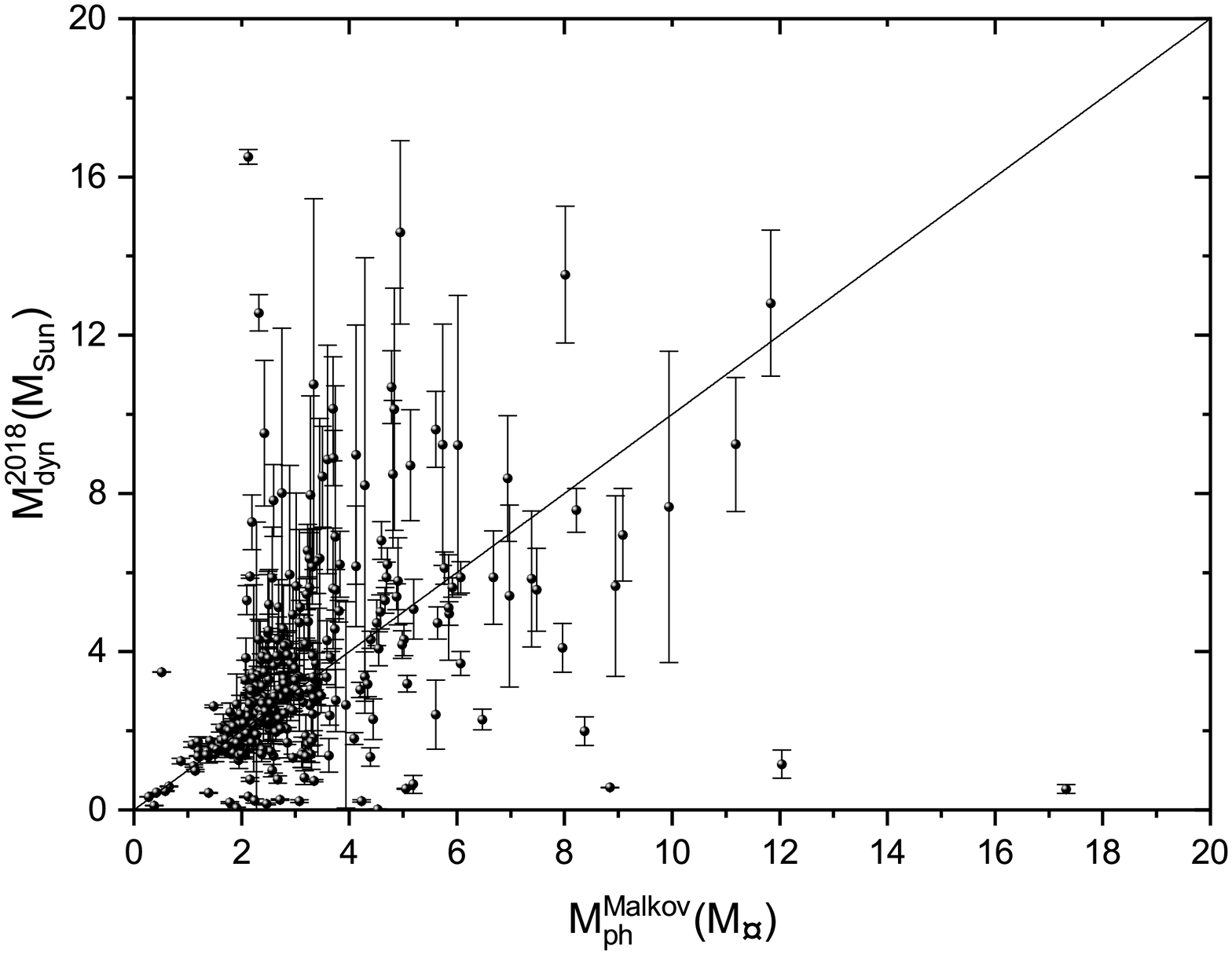}
  \caption{Dynamical mass depending on Gaia 2018 parallax measurements vs. Malkov photometric mass.}\label{egaia2018}
\endminipage
\end{figure}

The overlap of the BS studied by Malkov that have parallax measurements in the three catalogues is 340 CVBSs.

We plot here Malkov's photometric mass sums against the dynamical mass sums based on Hip 1997 parallax measurements in Fig.~\ref{ehip 1997}, Hip 2007 in Fig.~\ref{ehip 2007}, and Gaia DR2 2018 in Fig.~\ref{egaia2018}. The black lines represent the perfect fit $y=x$.
The figures show that the dynamical mass sums based on the three parallaxes are consistent with Malkov photometric masses at low masses, especially in the range ($0< \mathcal{M}\le 2$) $\mathcal{M}_{\odot}$. While there are clear discrepancies  for some more massive stars ($ \mathcal{M}\ge 2$) $\mathcal{M}_{\odot}$ better agreement is found for the measurements based on Hip 2007. However, this is not surprising because  Malkov used the Hip 2007 parallax measurements in his calculations of the individual photometric masses.

Any significant discrepancies may be related to mis-identification of stellar multiplicity for massive stars, where they are identified as binary systems but potentially have more than two components.  \cite{duchene2013stellar} pointed out that the probability of multiplicity of stars with $\mathcal{M}$ $\sim$ ($1.5\le \mathcal{M}\le 8$) $\mathcal{M}_{\odot}$ is $\ge 50$\% and for stars with $\mathcal{M}$ $\sim$ ($8\le \mathcal{M}\le 16$) $\mathcal{M}_{\odot}$ is $\ge 60$\%. Analysis of BSs using Al-Wardat's method can test this idea.

\subsection{Masses Based on Al-Wardat's Method}

Al-Wardat's method is a computational spectrophotometric multi-parameter approach that employs atmosphere modelling (ATLAS9) and synthetic photometry to estimate all physical parameters including individual masses (Note that Gaia team used ATLAS9  synthetic spectral library for the extinction and reddening estimations).
Since it depends on accurately measured magnitudes and colour indices, the method can estimate  masses and  parallaxes for MS stars without parallax information. However, the approach is more robust when parallaxes are available, especially for evolved stars.
We list here, in Table~\ref{mass alwardat}, the masses of 17 BSs  analyzed earlier using Al-Wardat's method, in addition to the results of the BS Hip 68170 in this paper.
The individual and total masses of those systems are given in columns 1,2 and 3, their photometric mass sum as given by Malkov are column 4 along with their dynamical mass sums and  uncertainties based on Hip 1997, Hip 2007, and Gaia 2018 parallax measurements in columns 5-10.
The authors using Al-Wardat's method employed the trigonometric parallax of Hip 1997 to analyze the systems Hip (83064, 83791, 11352, 11253, 4809),  Hip 2007 to analyze the systems Hip (70973, 72479, 689, 17491, 95995, 12552, 64838, 105947), and  Gaia DR2 to analyze the systems  Hip (14230, 14075, 116259).

\begin{figure}
\minipage {0.4\textwidth}
  \includegraphics[width=\linewidth]{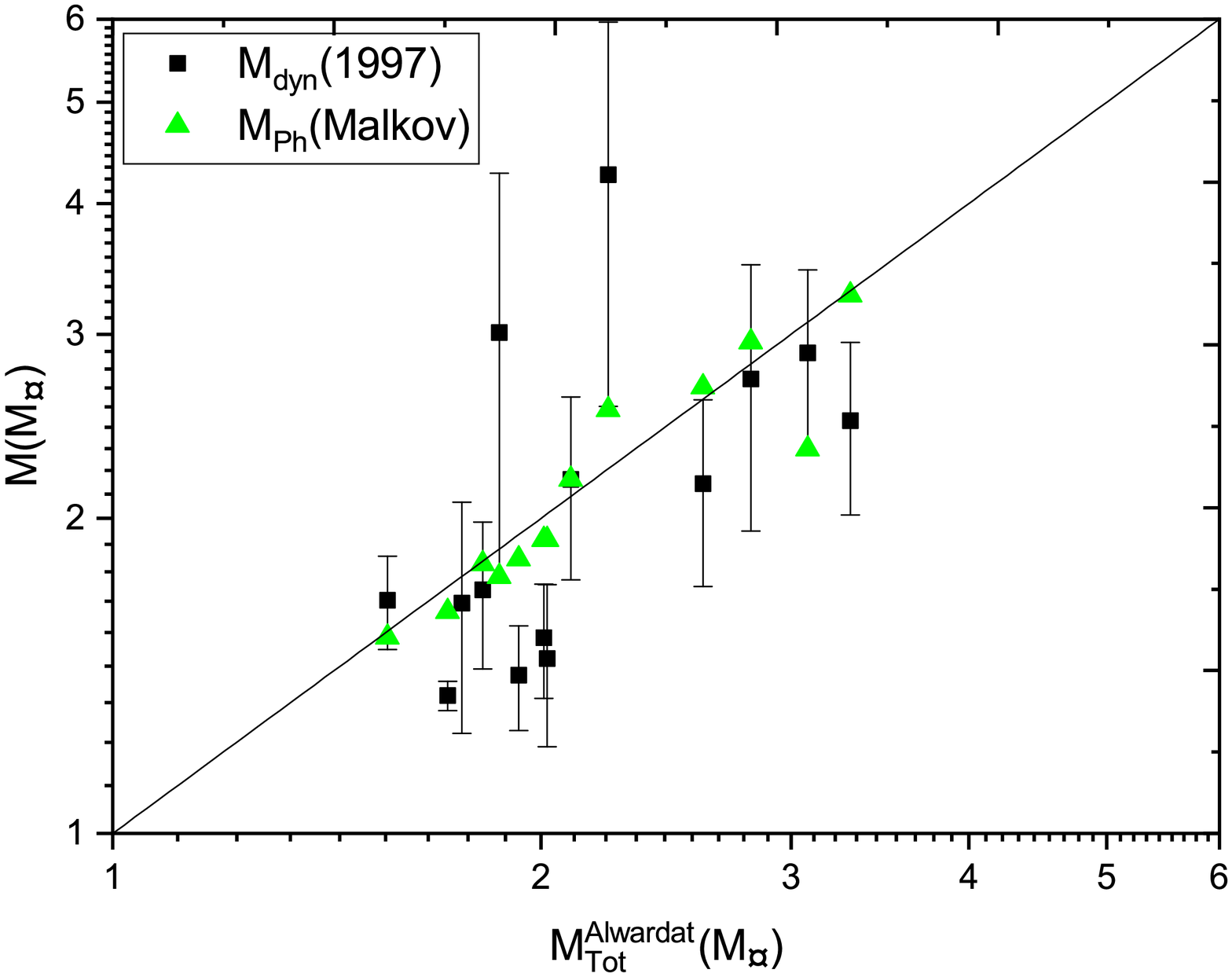}
  \caption{Malkov's photometric mass and dynamical mass based on Hipparcos 1997 parallax measurements vs. Al-Wardat's mass sum.} \label{alw97}
\endminipage\hfill
\minipage{0.4\textwidth}
  \includegraphics[width=\linewidth]{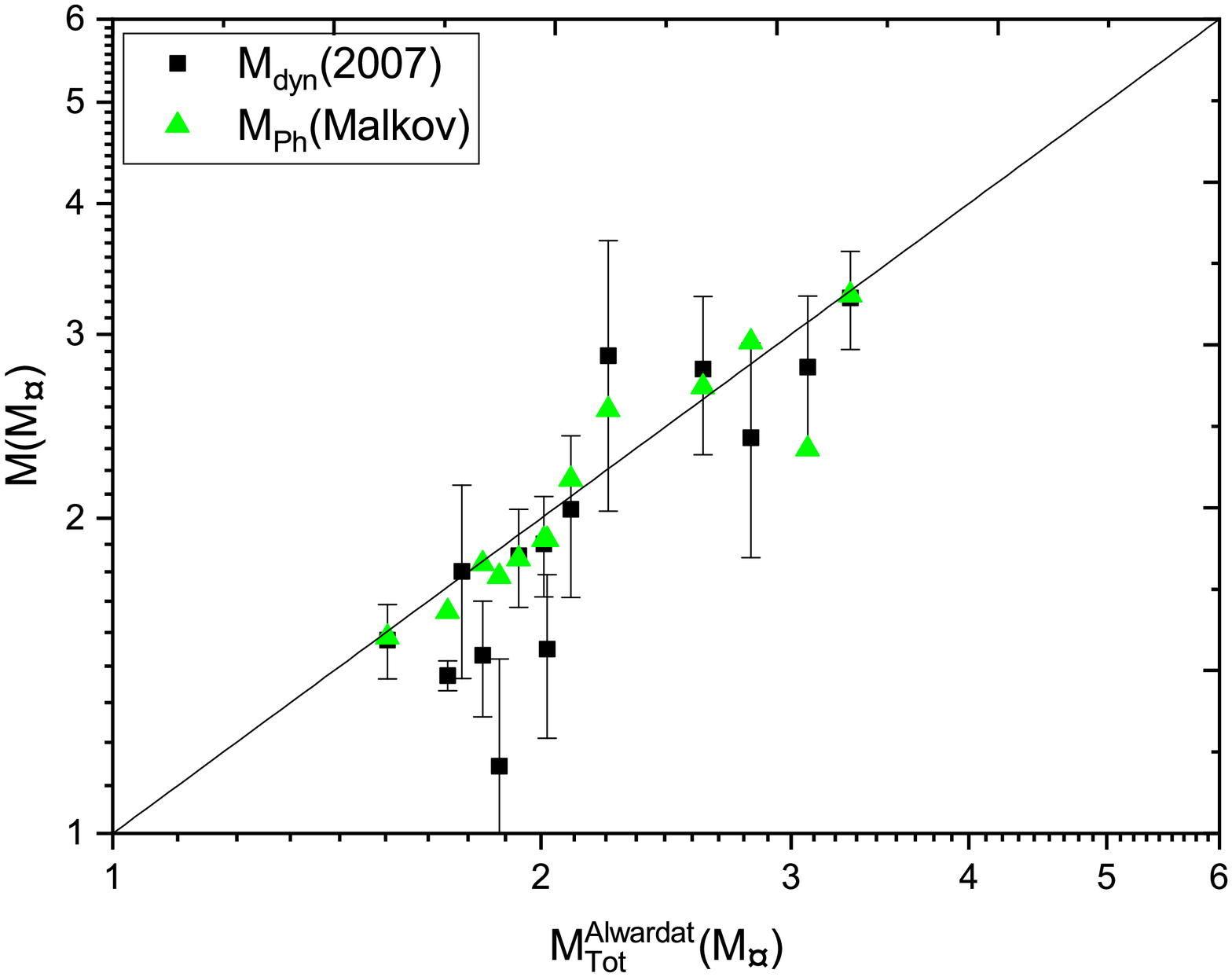}
  \caption{Malkov's photometric mass and dynamical mass based on Hipparcos 2007 parallax measurements vs. Al-Wardat's mass sum.} \label{alw07}
\endminipage\hfill
\minipage{0.4\textwidth}%
  \includegraphics[width=\linewidth]{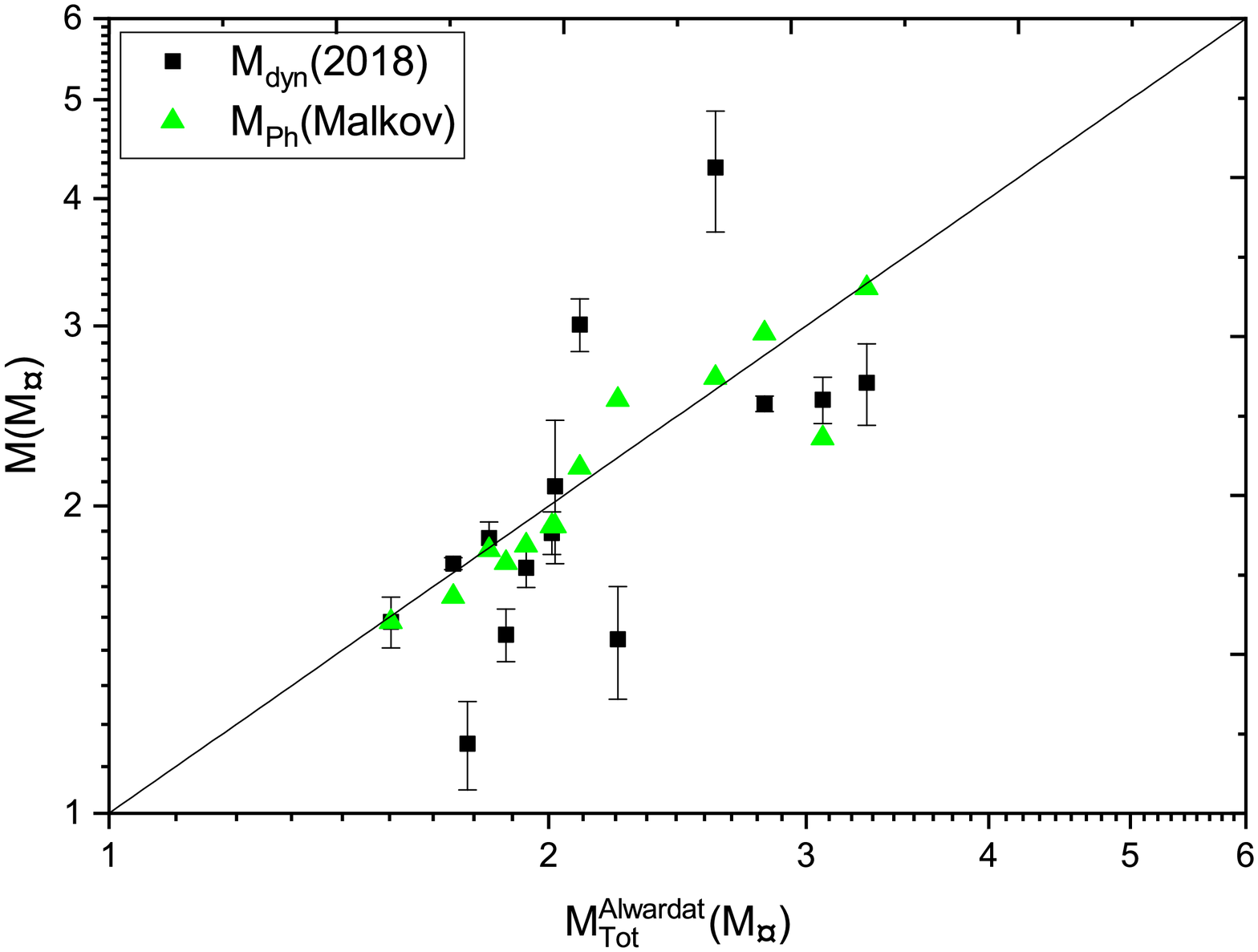}
  \caption{ Malkov's photometric mass and dynamical masses based on Gaia DR2 2018 parallax measurements vs. Al-Wardat's mass sum.} \label{alw18}
\endminipage
\end{figure}

We compare the mass sums estimated using  Al-Wardat's method with Malkov's photometric mass and the dynamical masses based on Hip 1997 parallax measurements in Fig.~\ref{alw97}, Hip 2007 in Fig.~\ref{alw07} and Gaia 2018 in Fig.~\ref{alw18}, where the black lines are the perfect fit $y=x$.
The comparison shows a very good consistency between Al-Wardat's masses and Malkov's photometric masses, and a good consistency between Al-Wardat's masses and the dynamical masses for most stars, although there are a few outliers in each figure. The best consistency is with the dynamical mass sums based on Hip 2007, but most of the stars were analyzed using the parallaxes from this catalogue.
We recommend that all systems should be reanalysed using  Gaia DR2 parallax measurements, which will be done in a future work.

\begin{table*}
	\centering
	\caption{The individual and total masses from AL-Wardat, Malkov photometric mass sum and dynamical mass sum.}
	\label{mass alwardat}
	\begin{tabular}{lcccccccccr}
		\hline
		Hip  & $\mathcal{M}_{1}$ & $\mathcal{M}_{2}$ & $\mathcal{M}_{Tot}$  & $\mathcal{M}_{Ph}$& $\mathcal{M}_{dyn}^{1997}$ & $\sigma \mathcal{M}$ & $\mathcal{M}_{dyn}^{2007}$ & $\sigma \mathcal{M}$ & $\mathcal{M}_{dyn}^{2018}$ & $\sigma \mathcal{M}$ \\
		
		References &$\mathcal{M}_{\odot}$ &$\mathcal{M}_{\odot}$ & $\mathcal{M}_{\odot}$& $\mathcal{M}_{\odot}$&$\mathcal{M}_{\odot}$ &$\mathcal{M}_{\odot}$ &$\mathcal{M}_{\odot}$& $\mathcal{M}_{\odot}$&$\mathcal{M}_{\odot}$& $\mathcal{M}_{\odot}$ \\
		\hline
	       83064 & 1.55 & 1.26 &	2.81&2.95 & 2.72& 0.77& 2.39& 0.55&	2.52& 0.04\\
	       \cite{2007AN....328...63A}& & & & & & & & &\\
	       83791& 1.33 & 1.13 & 2.46& 2.43& 2.36& 0.58&2.39& 0.58 & \space &\space\\
	       \cite{2007AN....328...63A}& & & & & & & & & \\
	       11352 & 0.93 & 0.89 & 1.82& 1.81 & 1.71 & 0.27 & 1.48 & 0.18& 1.86 & 0.07\\
	       \cite{2009AN....330..385A}& & & & & & & & &\\
	       11253& 1.06 & 0.7& 1.76 & &1.66& 0.41& 1.78& 0.37&	1.17& 0.12\\
	       \cite{2009AstBu..64..365A}& & & & & & & & & \\
	       70973& 1.07& 0.94& 2.01& 1.91 &1.54& 0.19& 1.89& 0.21& 1.88& 0.09\\
	       \cite{2012PASA...29..523A}& & & & & & & & &\\
	       72479&	0.94&	0.85&1.79&  & 1.28& 0.21& 1.58& 0.26 &\space&\space\\
	       \cite{2012PASA...29..523A}& & & & & & & & &\\
	       4809& 1.6& 1.46& 3.08& 2.33 &2.88& 0.57& 2.79& 0.47& 2.54& 0.132 \\ \cite{2013IJMPS..23...64A}& & & & & & & & &\\
	       689& 1.35& 1.25& 2.6& 2.67& 2.16&	0.44& 2.78& 0.48& 4.29& 0.58\\ \cite{2014PASA...31....5A}& & & & & & & & &\\
	       17491& 0.71& 0.86& 1.56& 1.54&1.67& 0.17& 1.53& 0.12& 1.54& 0.09 \\ \cite{2014AstBu..69..198A} & & & & & & & & &\\
          95995& 0.89& 0.83& 1.72& 1.63&1.35& 0.04& 1.41& 0.05&	1.75& 0.02\\ \cite{2016RAA....16..112M} & & & & & & & & &\\
          12552& 1.17& 1.06& 2.23&2.54& 4.26& 1.70& 2.86& 0.83& 1.48& 0.18 \\ \cite{2016RAA....16..166A} & & & & & & & & &\\
          64838& 1.75& 1.55& 3.3&3.27& 2.48& 0.46& 3.25& 0.35& 2.64& 0.24\\ \cite{2017AstBu..72...24A} & & & & & & & & &\\
          105947& 1.21& 0.89& 2.1& 2.18&2.18& 0.43& 2.04& 0.36& 3.01& 0.18\\  \cite{2018RAA....18...72M}& & & & & & & & & \\
          14230& 1.18& 0.84& 2.02&1.91& 1.47& 0.26& 1.5& 0.26& 2.09& 0.34\\ \cite{2018JApA...39...58M}& & & & & & & & & \\
          14075& 0.99& 0.877& 1.87&1.76& 3.01& 1.26& 1.16& 0.31& 1.49& 0.09\\ \cite{2018JApA...39...58M} & & & & & & & & &\\		
   	      116259& 1.18& 0.75& 1.93&1.83& 1.42& 0.16& 1.84& 0.19& 1.74& 0.08\\ \cite{2019RAA....19..105M} & & & & & & & & &\\
   	      	\hline	
   	      68170 & 1.49 & 1.46 & 2.95 &2.34 &  2.61 &  &  2.34 & & 13.66 & \\ This work & & & & & & & & &\\
			\hline
			HD 25811& 1.55& 1.5& 3.05& &\space& \space&\space& \space& 4.31 & -\\
			\cite{2014PASA...31....5A}& & & & & & & & &\\
			\hline
	\end{tabular}
\end{table*}

\section{Notes on specific systems}
\label{specific systems}

\begin{figure}
	\includegraphics[width=8cm]{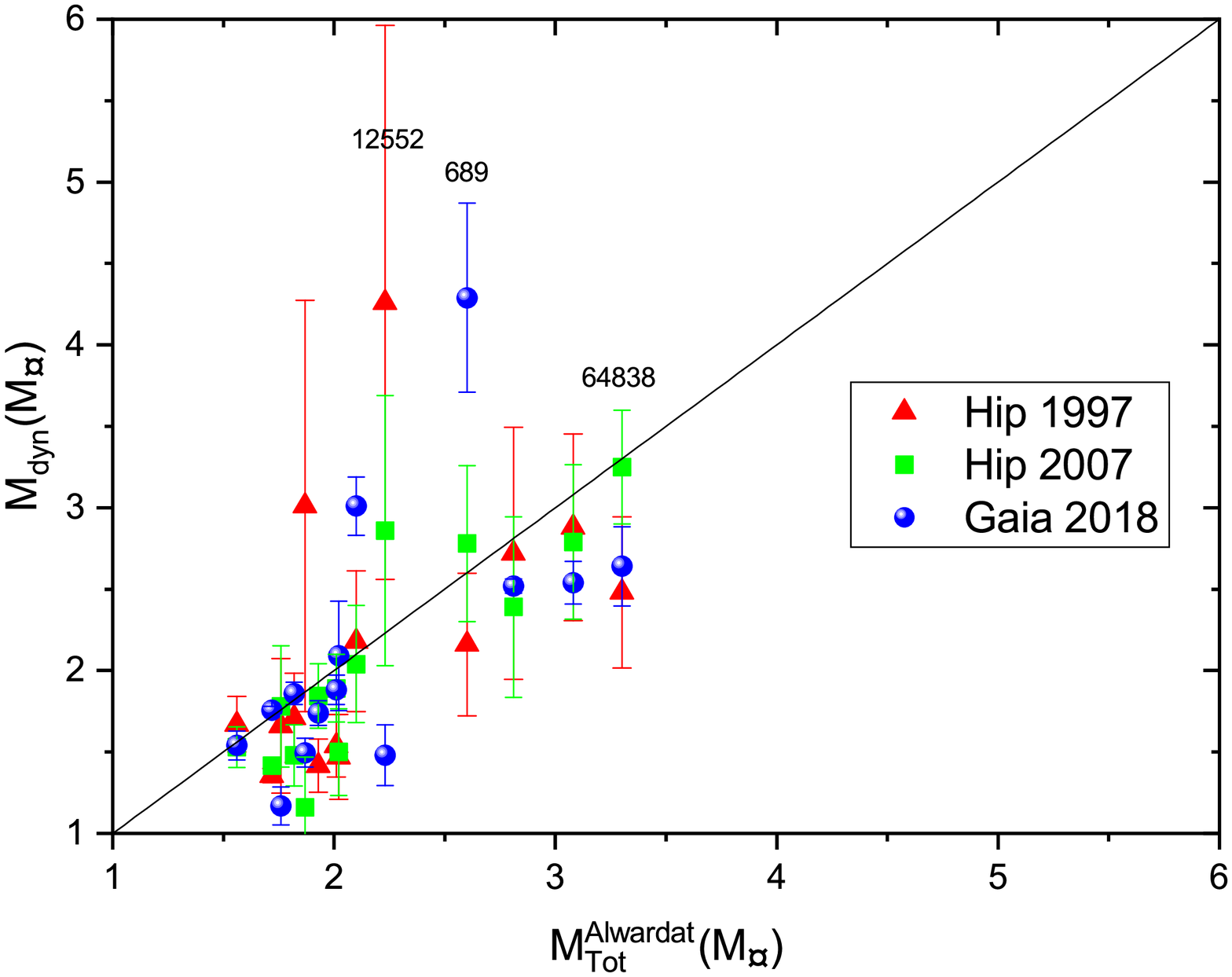}
    \caption{ The dynamical mass vs. Al-Wardat's mass sum.}
    \label{alwardat}
\end{figure}

Figs.~\ref{alw18} and \ref{alwardat} show some scattered points,  where there are significant differences between dynamical masses and those estimated by Al-Wardat's method. These are the systems HD 25811, Hip 12552, Hip 64838, and Hip 689. All of these have large error values for the  parallax measurements in the catalogues. A possible explanation is that the trigonometric parallax measurements have been distorted by the orbital motion of the components of such systems, which affects the position of the photo-center of the system ~\citep{1998AstL...24..673S}.
Individual systems are discussed below:

 HD 25811: This system was analyzed  using Al-Wardat's method ~\citep{2014PASA...31....5A}. In spite of the fact that there was no measured trigonometric parallax at that time, they estimated $M_a=1.55 \pm 0.16 $ $\mathcal{M}_{\odot}$, $M_b=1.50 \pm 0.15 $ $\mathcal{M}_{\odot}$, and a dynamical parallax ($\pi = 5.095 \pm 0.095$ mas, $d=196.27$ pc) depending on an initial  value  ($\pi = 5.24 \pm 0.6$ mas, $d=191$ pc) taken from \cite{2003PhDT.......174G}.
Comparing the estimated dynamical parallax by ~\cite{2014PASA...31....5A} as (5.095$\pm$ 0.095) mas with  the measured value by Gaia as (4.953$\pm$0.080) mas shows that the estimated value using Al-Wardat's method was very close to the new Gaia measurement. This is a good indicator of the accuracy of  the "Al-Wardat method" for analyzing CVBSs.
Another note regarding the system HD 25811 is that the published orbit in ORB6 is for ~\cite{balega2001infrared}, but two orbits were published after that; one by \cite{2003PhDT.......174G}, and the latest by ~\cite{2014PASA...31....5A}. We tried to modify the orbit using the new relative positional measurements in the Fourth Catalog of Interferometric Measurements of Binary Stars added to the system by Tokovinin \citep{2016AJ....151..153T}, but there was no difference in the solution, since the new point lies near the old points out of the clear arcs of the orbit.
Regarding the mass of the system,  ~\cite{balega2001infrared} proposed a mass sum of 4.31 $\mathcal{M}_{\odot}$, while \cite{2003PhDT.......174G} gave 3.76 $\mathcal{M}_{\odot}$, and ~\cite{2014PASA...31....5A} gave 3.05$\mathcal{M}_{\odot}$.

\begin{figure}
	\includegraphics[width=8cm]{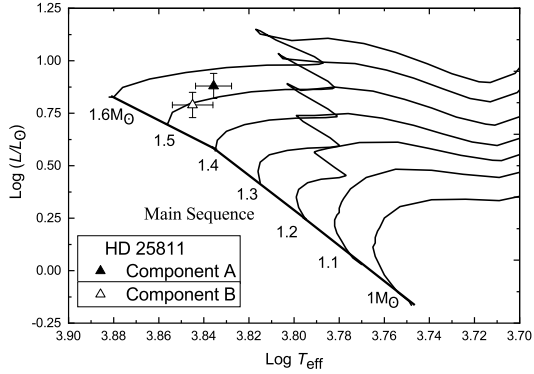}
    \caption{Positions of the components of HD 25811 on the evolutionary tracks of \citep{2000yCat..41410371G} for masses (1 , 1.1, ..., 1.6 $\rm\,M_\odot$).}
    \label{evohd}
\end{figure}

We reanalyzed the system using Al-Wardat's method and  the new Gaia trigonometric parallax. Fig \ref{evohd} shows the positions of the system's components on the evolutionary tracks of \cite{2000yCat..41410371G},    with a mass sum of $3.10\pm 0.37$ $\mathcal{M}_{\odot}$ which is very close to that of ~\cite{2014PASA...31....5A}. This is the closest mass to the 3.32 $\mathcal{M}_{\odot}$  calculated using  Gaia parallax  and the orbital elements of ~\cite{2014PASA...31....5A}, another indicator of the reliability of Al-Wardat's method. Adopting the  physical and geometrical parameters of the system given by ~\cite{2014PASA...31....5A} would suggest a  little bit higher parallax than that of Gaia, i.e the system is a bit closer.

 Hip 12552: This system has some discrepancies in trigonometric parallax, where  Hip 1997 gives 9.69 $\pm$ 1.29 mas,  Hip 2007 gives 11.07$\pm$ 1.07 mas and Gaia 2018 gives 13.786$\pm$ 0.583 mas. \cite{2016RAA....16..166A} estimated a value equal to 11.83 $\pm$ 1.07 mas, which is closest to Gaia measurement.  The discrepancy in the parallax measurements resulted in  differences in mass-sums, where the estimated mass-sum by \cite{2016RAA....16..166A}, based on Al-Wardat's method, was  2.23 $\mathcal{M}_\odot$ and the photometric mass sum was 2.54 $\mathcal{M}_\odot$, While this is  close to the dynamical mass based on Hip 2007 parallax measurements, we note again that it is mainly because Malkov used the parallax of Hip 2007. These mass sums are higher than the dynamical mass sum based on Gaia 2018 parallax measurements given as 1.48 $\mathcal{M}_\odot$.
 Reanalyzing the system using Al-Wardat's method assured a mass of the system higher than 2 solar masses, implying that the discrepancy could be due to an error in Gaia parallax measurement  for this  system, or due to inaccurate orbital elements. New relative positional measurements are required to resolve the situation.

 Hip 64838: This system has lower discrepancies in trigonometric parallax, where parallax from Hip 1997 is 13.45 mas, from Hip 2007 is 12.28 mas, and from Gaia 2018 is 13.18 mas. The photometric mass sum is 3.27 $\mathcal{M}_\odot$, identical to the dynamical mass based on Hip 2007 parallax measurements. However, the dynamical mass based on Gaia 2018 parallax measurements is 2.64 $\mathcal{M}_\odot$!.
  \cite{2017AstBu..72...24A} analyzed this system using both Al-Wardat's method and Docobo's dynamical method. The authors presented two orbital solutions; a short one with a period of 9.130 $\pm$ 0.030 yrs and a long one with a period of 18.442 $\pm$ 0.200 yrs, and two evolutionary states; either MS components or Sub-giant ones. The one preferred by the authors was the short period sub-giant solution, which required a dynamical parallax of 13.13 $\pm$ 0.43 mas and  a mass sum of 2.665 $\pm$ 0.125$\mathcal{M}_\odot$. This coincides perfectly with the trigonometric parallax  given later by Gaia and the masses calculated using it.

 Hip 689: This system was analyzed by ~\cite{2014AstBu..69..198A} using Al-Wardat's method asking the question: Is it a sub-giant binary?
 %based on Gaia parallax measurements, the system is a sub-giant binary.
The system has discrepancies in trigonometric parallaxes, where that of Hip 1997 is 12.72 mas, that of Hip 2007 is 11.69 mas, and that of Gaia 2018 is 10.112 mas. This gives a distance error range equals to 20 pcs. Hence, it affects the calculated mass-sums strongly.
Both the mass-sum given by  ~\cite{2014AstBu..69..198A} as 2.6 $\mathcal{M}_\odot$, and the photometric mass sum given by \cite{2012yCat..35460069M} as 2.67 $\mathcal{M}_\odot$ coincide with the  dynamical mass calculated based on Hip 2007 parallax measurement. However, when we recalculate the dynamical mass based on Gaia 2018 parallax measurement we obtain a value of 4.29 $\mathcal{M}_\odot$!. If we suppose that the parallax given by Gaia 2018 is precise, then we can say that Hip 689 has more than two components, and it could be a triple system. This requires a further analysis of the system and more high resolution imaging observations to resolve the question of multiplicity.

Hip 68170: This system is analyzed using Al-Wardat's method for the first time in this paper. It has been chosen as an example of the 55 problematic systems which have been discussed earlier as an example of the ability of the Al-Wardat's method to estimate the fundamental parameters independently of the parallax and to judge between different measurement methods.
The observational data used to analyze the system is collected in Table~\ref{observ}.
There are  clear discrepancies in the trigonometric parallaxes between Hip 1997 (14 mas),  Hip 2007 (14.43 mas), and  Gaia 2018 (8.06 mas). This gives a potential range in the possible distance of 50 pc. Hence, there is a corresponding large range of calculated dynamical mass-sums. The system also has a  very large value for the interstellar extinction of the system as 32.7542 as shown in Table~\ref{observ}.

\begin{table*}
	\centering
	\caption{Observational data of HIP 68170.} \label{observ}
	\begin{tabular}{ccc}\hline\hline\	Property &
	Hip 68170 & Source of data \\
	 & HD 121454   &\\
		\hline
		
		$\alpha_{2000}$  &  $13^ h$ $57^ m$  $21^{\rm s}32$ &  \cite{2001AJ....122.3480H}\\
		$\delta_{2000}$ & $-62^\circ 29' 20.''22$ & -\\
		
		$E(B-V)$ &
		$0.002\pm0.015$ &\cite{lallement20183d} \\
		$A_V$       &  0.0062 &  \cite{lallement20183d} 		\\
		$m_V$       &  6.65  & \cite{1997yCat.1239....0E}\\
		$m_B$       &  7.44  & -- \\ 		
		$(B-V)$ &   $0.79\pm0.004$  & --	\\
		$\pi_{Hip 1997}$ (mas)      & $14.00\pm0.79$ &  --	\\
		$\pi_{Hip 2007}$ (mas)    &  $14.43\pm0.65$  & \cite{van2007validation} \\
		$\pi_{Gaia 2018}$ (mas)      & $8.059\pm0.923$&\cite{2018yCat.1345....0G}\\
		$B_T$ & $7.637\pm0.005$ & \cite{2000AA...355L..27H}	\\			
		$V_T$  & $6.742\pm0.004$ & --\\
		$\triangle m_v$ $^*$& 0.39667 & $^{**}$  \\
		\hline\hline
	\end{tabular}\\
	$^*$ The average visual magnitude difference.\\
	$^{**}$ \cite{1997yCat.1239....0E, 2010AJ....139..743T, 2014AJ....147..123T, 2016AJ....151..153T, 2017AJ....154..110T}
	\\

\end{table*}

The results of the analysis for the three parallax measurements of Hipparcos and Gaia are listed in Table ~\ref{phy68170}, which gives the physical and geometrical parameters of the individual components (effective temperature, radii, gravity, luminosity's, absolute and bolometric magnitudes, spectral types, and masses).
The estimated mass sum using Al-Wardat's method are: 2.95 $\mathcal{M}_\odot$ using Hip 1997 parallax measurement, 2.91 $\mathcal{M}_\odot$ using  Hip 2007 parallax measurement and 4.07 $\mathcal{M}_\odot$ using Gaia DR2 parallax measurement.

In order to calculate the dynamical mass sum, we used the latest modified orbit of the system which gives an orbital period of $P=18.757$ years and semi-major axis of $a=0.136 arc-second$ \citep{2019yCat..51560240M}. Hence,
the dynamical mass sum is 2.61 $\mathcal{M}_\odot$ using Hip 1997 parallax measurement, 2.38 $\mathcal{M}_\odot$ using Hip 2007, and 13.66 $\mathcal{M}_\odot$ using Gaia DR2.

This gives a difference between Al-Wardat's estimated mass sum  and the dynamical mass sum  $\mathcal{M}_{dyn}-\mathcal{M}_{Tot}$ as: $\mathcal{M}_{dyn}-\mathcal{M}_{Tot}=0.34$ for Hip 1997, $\mathcal{M}_{dyn}-\mathcal{M}_{Tot}=0.53$ for Hip 2007, and  $\mathcal{M}_{dyn}-\mathcal{M}_{Tot}=9.59$ for Gaia DR2.
Which shows that  the discrepancy in the  parallax does not significantly affect the estimated masses using Al-Wardat's method, while it does have a clear impact on the dynamical mass sum.

\begin{figure}
	\includegraphics[width=8cm]{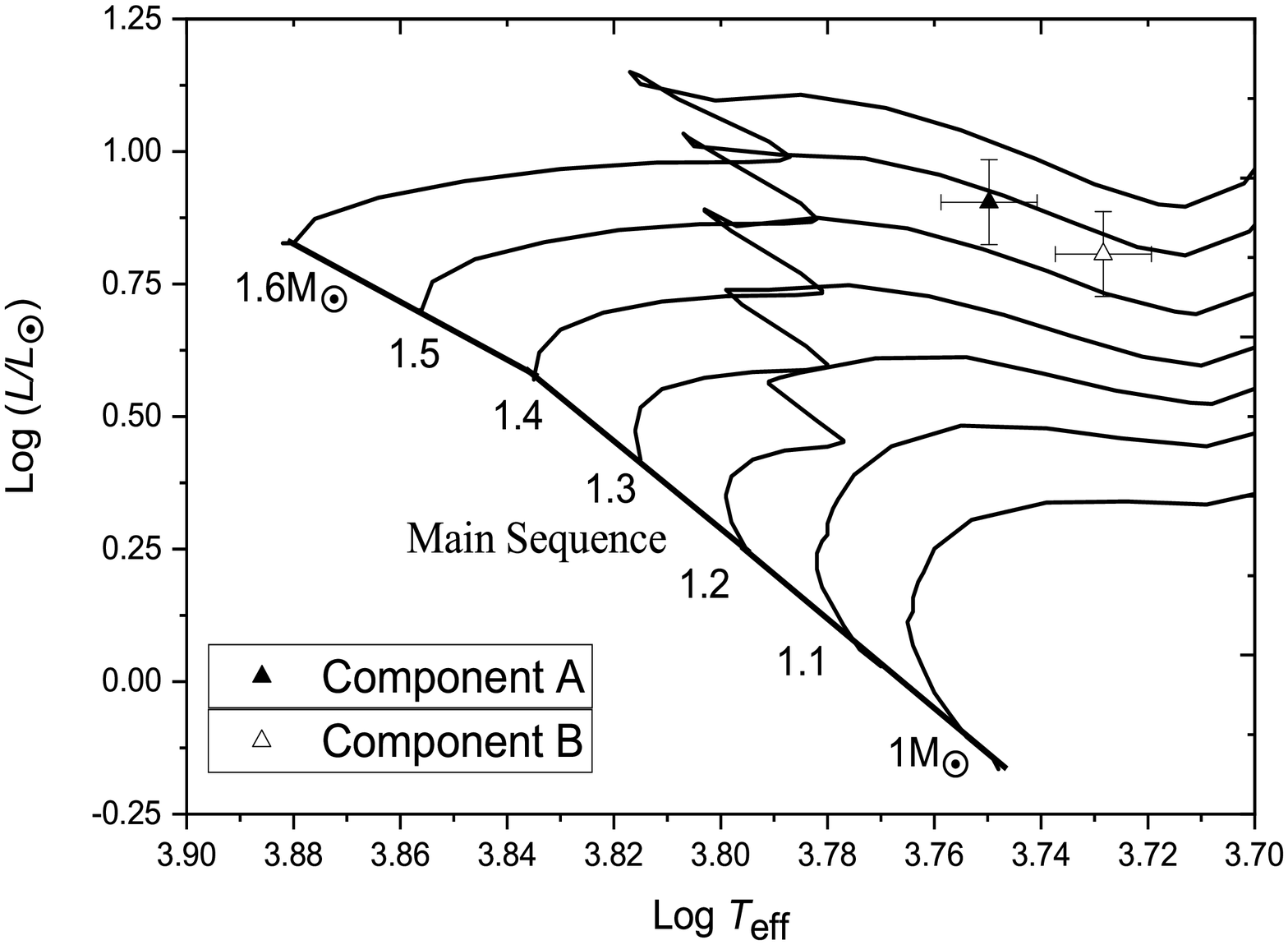}
    \caption{ Positions of the components of Hip 68170 on the H-R diagram based on the analysis using Al-Wardat's method. The evolutionary tracks are taken from ~\citep{2000yCat..41410371G} for masses (1 , 1.1, ..., 1.6 $\rm\,M_\odot$).}
    \label{ev07}
\end{figure}

So, the  final result for the system is as follows:  Depending on Fig. ~\ref{ev07}, Al-Wardat's method gives a mass sum of $2.95\pm1.12$ , this leads for a new dynamical parallax of $13.43\pm1.37$, to which the closest is that of Hip 1997 as $14.00\pm0.79$. This shows that there is a clear issue in the parallax measurements of Gaia for this system, likely to be mainly due to interstellar extinction.

%We   adopt  the solution that is calculated from the parallax of Hip 1997, as it is more precise than that of Hip 2007, and gives much better consistency between masses comparing with Gaia.

\begin{figure}
	\includegraphics[width=8cm]{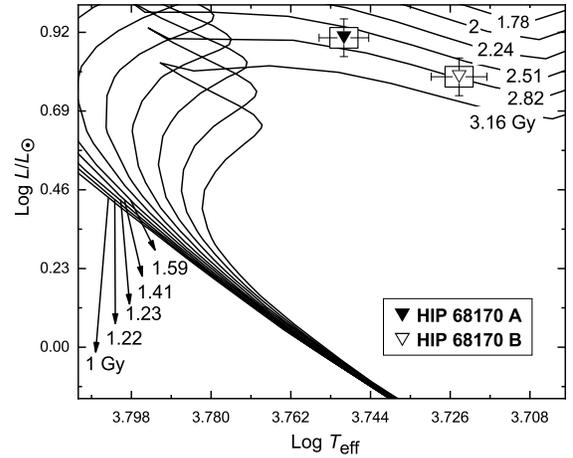}
    \caption{The isochrones for both components of HIP 68170 on the H-R diagram for low- and intermediate-mass stars, with the composition of [Z=0.019, Y=0.273]. The isochrones are taken from ~\citep{2000yCat..41410371G}.}
    \label{isochron 68}
\end{figure}

The analysis shows that the system consists of two subgiant stars, as shown in Fig. ~\ref{ev07}, with a metallicity of 0.019 dex and age of 2.75$\pm$0.50 Gy as shown in Fig.~\ref{isochron 68}. Fragmentation is the most probable formation theory for such a system.
The spectral types of the components are estimated as G6.5IV and G9.5IV for the primary and secondary components respectively, which are consistent with those proposed by \cite{2002A&A...384..491C} as G2IV/III and G4IV/III.

\begin{table*}
\centering
	\small
	\begin{center}
		\caption{The physical parameters of the individual components of the system  HIP\,68170 as estimated using Al-Wardat's method and based on parallax measurements of Hip 1997, Hip2007, and Gaia DR2. The adopted final results for the system are given in cols. 3 and 4, with a new dynamical parallax of 13.43 mas.}
		\label{phy68170}
		\begin{tabular}{cccc|cc|cc}
			\noalign{\smallskip}
			\hline\hline
			&
			&\multicolumn{6}{c}{HIP 68170}\\
			\cline{3-8}
			\noalign{\smallskip}
			&  &  \multicolumn{2}{c}{Using  $\pi_{Hip 1997}^{**}$} &  \multicolumn{2}{c}{Using	$\pi_{Hip 2007}$} &  \multicolumn{2}{c}{Using  $\pi_{Gaia 2018}$}\\
			\cline{3-8}
			\noalign{\smallskip}
			&  &  \multicolumn{2}{c}{$14.00\pm0.79$ (mas)} &  \multicolumn{2}{c}{ $14.43\pm0.65$(mas)}&  \multicolumn{2}{c}{       $8.059\pm0.923$(mas)}\\
			\cline{3-4}
			\cline{5-6}
			\cline{7-8}
		\noalign{\smallskip}
			Parameters & Units	&  A & B&  A & B&  A & B  \\
			\hline
			\noalign{\smallskip}
			$ T_{\rm eff}$ {$\pm$ $\sigma_{\rm T_{\rm eff}}$}& [~K~] & $5620\pm100$ & $5300\pm100$ & $5620\pm100$ & $5300\pm100$ & $5620\pm100$ & $5300\pm100$  \\
			
			R {$\pm$ $\sigma_{ R}$}  & [R$_{\odot}$] & $2.993\pm0.06$& $2.951\pm0.05$& $2.892\pm0.06$&$2.869\pm0.05$ &$5.221\pm0.06$&$5.124\pm0.05$\\
			$\log\rm g$ {$\pm$ $\sigma_{\rm log g}$} & [$cm/s^2$] & $3.65\pm0.11$ & $3.66\pm0.13$  & $3.67\pm0.11$ &$3.67\pm0.13$ & $3.30\pm0.11$ &$3.31\pm0.13$  \\
			
			$ L $ {$\pm$ $\sigma_{L}$} & [$ L_\odot$] &$8.02\pm0.30 $  &$6.17\pm0.10 $   &$7.49\pm0.30 $  &$5.83\pm0.10 $ &$24.41\pm0.30 $  &$18.59\pm0.10 $ \\
			$M_{bol}$ {$\pm$ $\sigma_{ M_{bol}}$} & [mag] & $2.54\pm0.08$ &$2.82\pm0.08$ & $2.61\pm0.08$ &$2.88\pm0.08$ & $1.33\pm0.08$ &$1.62\pm0.08$\\
			$ M_{V}$ {$\pm$ $\sigma_{ M_{V}}$} & [mag] & $2.63\pm0.13$ & $2.92\pm0.14$& $2.70\pm0.13$ & $2.99\pm0.14$  & $1.50\pm0.13$ & $1.77\pm0.14$ \\
			
			$\mathcal{M}$ 	& [$\mathcal{M}_{\odot}$] &$1.49\pm0.16$&1.46$\pm1.46$ &$1.47\pm0.16$&$1.44\pm0.15$ &$2.05\pm0.16$&$2.02\pm0.15$\\
			
			Sp. Type$^{*}$ & &  G6.5 IV & G9.5 IV  &  G6.5 IV & G9.5 IV   &  G6.5 IV & G9.5 IV\\
			
		$\pi_{dyn}$ & [mas]&  \multicolumn{2}{c}{$13.43\pm 1.12$}  & \multicolumn{2}{c}{$13.49$}&\multicolumn{2}{c}{$12.06$} \\
			\hline\hline
			\noalign{\smallskip}
		\end{tabular}\\
	%	$^{*}${Based on the evolutionary tracks of~\citep{2000yCat..41410371G}.\\}
		$^{*}${Based on temperature-Sp. type from Lang tables \cite{1992adps.book.....L}. \\}
		$^{**}${We adopt this solution. \\}
	\end{center}

\end{table*}

\section{Discussion}

A comparison between the calculated masses using trigonometric parallaxes of the three  catalogues has shown a consistency in the distributions.  Of course, this is an expected result since the dynamical masses in the three cases are calculated using the same orbital solutions, in addition to the fact that the sample focused on systems with differences between the three parallaxes  within 5\% of the value. Systems of higher values and larger differences in parallax measurements between Gaia and Hipparcos (both catalogues),  are eliminated from this study, and will be studied again after the next release of Gaia data.
The distribution of the masses shows a concentration of BSs among the low mass sums $\mathcal{M}$ $\sim$ ($0< \mathcal{M}\le 4$) $\mathcal{M}_{\odot}$, which is reasonable because stars in the milky way galaxy are mainly  main-sequence stars  with  masses in the range $\mathcal{M}$ $\sim$ ($0.08 < \mathcal{M}\le 8$) $\mathcal{M}_{\odot}$.

This tendency for the binaries to have low masses could be explained by the theories of  formation of BS, but these still need more observational data to differentiate between one theory or another  \citep{2018AJ....156...48T}. In general, the currently most accepted theory of the formation of BS is the fragmentation of proto-stellar cores or circumstellar discs \citep{1995MNRAS.277..362B, 1995MNRAS.277.1491K, 2002MNRAS.336..705B, 2002ARA&A..40..349T, 2006MNRAS.373.1563K, clarke2009pseudo, offner2010formation, 2016ARA&A..54..271K, moe2017mind, 2019ApJ...875...61M, 2020MNRAS.491.5158T}. Which may result in multiple systems higher than binaries in the case of  massive proto-stellar cores and discs, because of the gradually increasing likelihood that more massive stars will fragment \citep{2006MNRAS.373.1563K}.
We expect a strong contribution from Gaia data in solving the mysteries of formation of multiple stars, where recent stellar evolution theories concentrate on the study of massive stars \citep{2020MNRAS.492.2497A}.

 The rich data of Gaia DR2 are already being implemented in solving orbits and masses of binary and multiple systems, see for example Tokovinin's work  ~\citep{ 2018AJ....156..194T, 2018AJ....156...48T, 2019yCat..51560048T, 2019AJ....157...91T, 2019AJ....158..167T, 2019AJ....158..222T}. It is also very useful in determining the physical and geometrical parameters of binary stars, using Al-Wardat's method, which has been used to analyze several solar type binary and triple systems ~\citep{2009AN....330..385A, 2009AstBu..64..365A, 2012PASA...29..523A, 2014AstBu..69..198A,2016RAA....16..166A,2016RAA....16..112M,2017AstBu..72...24A, 2019AstBu..74..464M}, and was successfully  applied to sub-giant BSs like   HD\,25811,  HD\,375, HD\,6009 ~\citep{2014PASA...31....5A,2014AstBu..69...58A, 2014AstBu..69..454A} and Hip\,68170 in this work.

The accuracy of the Al-Wardat method had been demonstrated in this paper shown to provide a useful consistency check for DR2 parallax measurements. Some of the parallaxes obtained by Hipparcos give masses more consistent with the photometric and dynamic system parameters than DR2. On the other hand, Al-Wardat's method estimated a parallax for the system HD25811, which had no parallax from Hipparcos, as 5.095 $\pm$0.095 mas in 2014 ~\citep{2014PASA...31....5A}. This value is very close to that of Gaia DR2 - 4.953 $\pm$ 0.081 mas.
 Moreover, the method can deal with multiple stellar systems which are sometimes ignored by dynamical methods and the Malkov Method, which assumes that the systems are  binaries.

\section{CONCLUSION}

In 2018, the Gaia collaboration released the DR2 data, which gave precise parallax measurements for approximately 1.7 billion objects in addition to other photometric and astrometric data.
These precise parallax measurements have allowed many astronomical questions to be addressed. One in particular is the case of close visual binary stars, where it has been noted that Hipparcos parallax measurements of binary and multiple systems are, in some cases, distorted by the orbital motion of the components of such systems ~\citep{1998AstL...24..673S}.
In this paper, we have looked at the precision of the parallax measurements for the two missions and considered how they affect the measurements of the physical parameters for a sample of 1700 close visual binaries, taken from the  Sixth Catalog of Orbits of Visual Binary Stars.
First, we focused on comparing the parallax measurements between the three space-based astrometric catalogues: Hipparcos 1997, van Leeuwen 2007, and Gaia DR2 2018. The results showed that van Leeuwen's reduction of Hipparcos data was indeed an improvement on Hipparcos 1997, and those parallaxes are in better agreement with the Gaia DR2 release than those of Hipparcos 1997.
Secondly, this work studied the mass-sum of the selected binary systems, where we calculated the dynamical mass-sum using parallaxes from the three catalogues and then compared the results with masses estimated using other methods (340 systems from Malkov photometric masses and 17 systems depended on Al-Wardat's method).
The results showed that the estimated masses using Al-Wardat's method for analyzing CVBSs, which is a computational spectrophotometric technique, were closer to the dynamical masses than those of the photometric mass sum given by Malkov, and closer to the dynamical masses calculated using  van Leeuwen 2007 parallaxes. The latter point can be explained by noting that those works, which used Al-Wardat's method, adopted mainly van Leeuwen 2007 parallax measurements.
Finally, we discussed five specific BSs which showed discrepancies between their mass sums calculated or estimated by different methods. The comparison showed that Al-Wardat's method is an effective method for analyzing close visual binary an multiple systems.

There are several future lines of study that have emerged from our work:
\begin{itemize}
    \item Interstellar extinction should be taken into account during the further analysis of Gaia parallax measurements. Special attention should be given to specific high extinction regions in the galaxy.
    \item The effect of duplicity, multiplicity on the photo center and resulting Gaia parallax measurements should also be taken into account.
    \item There needs to be a detailed programme to reanalyze all previously studied binary and multiple systems using the new Gaia parallax measurements and applying the Al-Wardat Method for complete internal consistency in the measurement of the stellar physical parameters.
    \item A new parallax measurements for the system Hip\,12552 is needed, and also new relative position measurements to help in solving the issue of  the parallax difference between Gaia and Hipparcos, and to obtain the system's precise fundamental parameters. This may become available from the next Gaia data release.
    \item To reanalyze the system Hip\,689 using a different method in order to know if it is a binary, a triple or a quadruple system.
\end{itemize}

\begin{acknowledgements}
This work has made use of data from the European Space Agency (ESA) mission
{\it Gaia} (\url{https://www.cosmos.esa.int/gaia}), processed by the {\it Gaia}
Data Processing and Analysis Consortium (DPAC,
\url{https://www.cosmos.esa.int/web/gaia/dpac/consortium}).
It also has made use of SAO/NASA, the SIMBAD database, the Fourth Catalog of Interferometric Measurements of Binary Stars, IPAC data systems, and codes of Al-Wardat's method for analyzing close visual binary and multiple stars.

Funding for the DPAC
has been provided by national institutions, in particular the institutions
participating in the {\it Gaia} Multilateral Agreement.

MAB acknowledges the support of the UK Space Agency (grant numbers ST/S001123/1, ST/N0009978/1, PP/D006511/1).

\end{acknowledgements}

\section*{DATA AVAILABILITY}
The data underlying this article will be shared on reasonable request to the corresponding author.

%\bibliographystyle{pasa-mnras}
%\bibliography{1r_lamboo_notes}

\end{document}